# Physics-informed Neural Operator Learning for Nonlinear Grad-Shafranov Equation

Siqi Ding [1,2], Zitong Zhang [3,4], Guoyang Shi [1,2], Xingyu Li [5], Xiang Gu [1,2], Yanan Xu [1,2], Huasheng Xie [1,2], Hanyue Zhao [1,2], Yuejiang Shi [1,2], Tianyuan Liu [1,2]*

([1] *ENN Science and Technology Development Co., Ltd., Langfang, 065991, China*;

[2] *Hebei Key Laboratory of Compact Fusion, Langfang, 065001, China*;

[3] *College of Mechanical Engineering, Xi'an University of Science and Technology, Xi'an 710054, China*;

[4] *MOE Key Laboratory of Thermo-Fluid Science and Engineering, School of Energy and Power Engineering, Xi'an Jiaotong University, Xi'an 710049, China*;

[5] *School of Physics, Dalian University of Technology, Dalian 116024, China*)

*Corresponding author: liutianyuan@enn.cn

**Abstract**: As artificial intelligence emerges as a transformative enabler for fusion energy commercialization, fast and accurate solvers become increasingly critical. In magnetic confinement nuclear fusion, rapid and accurate solution of the Grad-Shafranov equation (GSE) is essential for real-time plasma control and analysis. Traditional numerical solvers achieve high precision but are computationally prohibitive, while data-driven surrogates infer quickly but fail to enforce physical laws and generalize poorly beyond training distributions. To address this challenge, we present a Physics-Informed Neural Operator (PINO) that directly learns the GSE solution operator, mapping shape parameters of last closed flux surface to equilibrium solutions for realistic nonlinear current profiles. Comprehensive benchmarking of five neural architectures identifies the novel Transformer-KAN (Kolmogorov-Arnold Network) Neural Operator (TKNO) as achieving highest accuracy (0.25% mean $L_2$ relative error) under supervised training (only data-driven). However, all data-driven models exhibit large physics residuals, indicating poor physical consistency. Our unsupervised training can reduce the residuals by nearly four orders of magnitude through embedding physics-based loss terms without labeled data. Critically, semi-supervised learning—integrating sparse labeled data (100 interior points) with physics constraints—achieves optimal balance: 0.48% interpolation error and the most robust extrapolation performance (4.76% error, 8.9× degradation factor vs 39.8× for supervised models). Accelerated by TensorRT optimization, our models enable millisecond-level inference, establishing PINO as a promising pathway for next-generation fusion control systems.

**Key words**: Nuclear Fusion; Physics-informed Neural Operator; Grad-Shafranov Equation

## 1 Introduction

In the quest for clean, sustainable fusion energy, magnetic confinement devices like tokamaks represent a leading approach. The convergence of artificial intelligence and fusion energy research marks a paradigm shift in the path toward commercial fusion power. Recent breakthroughs have demonstrated that machine learning can fundamentally accelerate fusion development by enabling capabilities previously unattainable with traditional methods: deep reinforcement learning has achieved autonomous magnetic control of tokamak plasmas [1] and active prevention of tearing instabilities [2], while novel neural state-space models have proven capable of mastering complex plasma dynamics to robustly navigate critical non-stationary phases like pulse ramp-down [3]; AI-driven optimization of 3D fields successfully suppresses edge energy bursts without compromising fusion performance [4]; multimodal super-resolution techniques are now unlocking hidden physics by reconstructing high-fidelity data from limited measurements [5]. Together with established AI disruption prediction systems [6], these advancements underpin the "AI-Fusion Digital Convergence" explicitly prioritized in the U.S. Department of Energy's 2025 Fusion Roadmap as a key enabler for commercial deployment by the mid-2030s [7].

Within this rapidly evolving landscape, the development of fast, accurate, and physically consistent solvers has become a critical bottleneck. This challenge is particularly acute for tokamak design, where stable operation and efficiency hinge on a precise understanding of magnetohydrodynamic (MHD) equilibrium. This equilibrium, where the plasma pressure gradient is exactly balanced by the Lorentz force ( $\nabla \mathbf{P} = \mathbf{J} \times \mathbf{B}$ ), serves as the theoretical foundation for nearly all key physics analyses. It provides the necessary magnetic field configuration for evaluating MHD instabilities that define the operational limits of a device and can lead to catastrophic disruptions. Moreover, since micro-turbulence—the primary driver of heat and particle transport—is strongly dependent on the macroscopic magnetic geometry, an accurate equilibrium solution is a prerequisite for high-fidelity transport simulations [8]. Experimentally, maintaining MHD equilibrium is fundamental throughout the entire discharge process. Precise control over specific equilibrium configurations is especially crucial for developing advanced operational scenarios like those with Internal Transport Barriers or in High-confinement mode (H-mode) [9].

Fundamentally, magnetohydrodynamics represents an extremely complex multi-scale, nonlinear physical phenomenon arising from the coupling of Maxwell's equations and the Navier-Stokes equations, however, making it a mathematically formidable problem to solve. In axisymmetric tokamaks, this complex 3D problem is elegantly reduced to a 2D nonlinear partial differential equation: the Grad-Shafranov (GS) equation [10, 11]. This equation is the linchpin connecting fusion theory, experimental control, and reactor design, making the development of accurate and fast GS solvers a critical task for the entire field.

The demand for GSE solvers creates a persistent trade-off between computational speed and accuracy, which has become a major bottleneck in both experimental operations and theoretical modeling. For real-time feedback control in tokamaks, equilibrium reconstruction must be performed on millisecond or even sub-millisecond timescales to adjust external coil currents and maintain the plasma's position and shape [12]. For instance, modern isoflux control loops can operate at several kilohertz, requiring a single equilibrium calculation to be completed within hundreds of microseconds [13]. On the other hand, in integrated modeling frameworks used for simulating entire discharge scenarios (e.g., OMFIT [14]), the GS solver is called thousands of times, coupling with transport, heating, and current drive modules [14]. The cumulative computational cost of the GS solver thus becomes the primary bottleneck, severely limiting the efficiency of scientific discovery and scenario optimization. This challenge extends to the design of next-generation devices like ITER and DEMO, where extensive parameter scans are necessary to explore the vast design space [15].

To address this challenge, various methods have been developed. Traditional approaches include analytical methods, which provide fast solutions by simplifying the plasma pressure and current profiles but lack accuracy for realistic scenarios [16-19], and numerical iterative solvers. Numerical solvers, based on finite difference method (FDM) or finite element method (FEM), offer high accuracy for complex, nonlinear profiles and are the cornerstone of fusion research. They are categorized as forward solvers (e.g., CHEASE [20]), which compute the equilibrium from given plasma profiles, and inverse solvers (e.g., EFIT [21]), which reconstruct the internal plasma state from external magnetic measurements. However, their iterative nature makes them computationally expensive and sometimes prone to convergence issues, rendering them unsuitable for real-time applications [22]. In recent years, machine learning methods, especially deep learning, have emerged as promising alternatives for building surrogate solvers [23, 24]. Under a supervised learning paradigm [17], models are trained on large datasets generated by conventional numerical solvers and can subsequently perform inference in milliseconds. However, their performance degrades sharply when encountering scenarios outside the training distribution—a “generalization crisis” that limits their use in exploring novel operational regimes.

Physics-Informed Machine Learning (PIML) addresses this limitation [25]. Physics-Informed Neural Networks (PINNs), a prominent PIML method, incorporate the residual of the governing PDE into the loss function, eliminating the need for exhaustive labeled datasets while enhancing physical consistency [25-27]. However, standard PINNs struggle with complex nonlinear problems and must be retrained for any change in equation parameters or boundary conditions [28, 29]. To overcome this, Neural Operators (NO) were introduced to learn the solution operator itself—a mapping from input parameter functions to solution functions. Architectures like the Deep Operator Network (DeepONet) [30] and the Physics-informed Neural Operator [28], which combines a NO with a PDE residual loss, represent the state-of-the-art for solving complex parameterized PDEs. These operator-based models have achieved significant breakthroughs in other scientific domains, such as weather forecasting, with models like FourCastNet [31] for ensemble forecasting and NeuralGCM [32] for climate simulation, both delivering predictions orders of magnitude faster than traditional numerical models while maintaining comparable accuracy.

The application of PIML to the GSE has shown initial promise across several problem formulations:

**Forward solvers (fixed boundary)**: Several studies have successfully used PINNs for fixed-boundary equilibria, primarily with simplified linear [27] or Solov’ev [26, 28] source terms. Zhou et al. [27] achieved high precision for single-parameter cases, while Rizqa et al. [28] benchmarked two architectures.

**Free-boundary solvers**: Wang et al. [29] explored free-boundary problems but without extrapolation or generalization testing. The added complexity of self-consistently determining the boundary shape remains a significant challenge.

**Inverse problem** (equilibrium reconstruction): Recent work has applied neural networks (NNs) to reconstruct internal plasma profiles from external magnetic measurements [33], though rigorous uncertainty quantification is lacking.

Despite these advances, the field still has significant gaps: (1) most research relies on simplified current profile models, neglecting the strong nonlinearities present in realistic plasma current profiles; (2) the crucial ability of models to extrapolate to new parameter regimes remains largely unverified; and (3) there is a lack of rigorous benchmarking that compares the performance of different advanced deep learning architectures, especially for parameter extrapolation and model generalization.

To situate this work, Table 1 compares recent NN-based GS solver studies, highlighting the research gaps we aim to address.

Table 1. Comparison of recent neural network-based GS solvers.

| Study | Nonlinear profiles | Boundary condition | Extrapolation analysis | Learning comparison | Architecture comparison |
|---|---|---|---|---|---|
| B. Jang et.al [34] | ✗ | Fixed | ☑ | ✗ | ✗ |
| F. N. Rizqan et.al [35] | ✗ | Fixed | ☑ | ✗ | ☑ |
| C. Zhou et.al [36] | ✗ | Fixed | ✗ | ✗ | ✗ |
| Z. Wang et.al [37] | ☑ | Free | ✗ | ✗ | ✗ |
| **Our work** | ☑ | **Fixed** | ☑ | ☑ | ☑ |

This work advances the state-of-the-art in surrogate modeling for tokamak equilibrium through three key contributions:

1. **First PINO Application to Nonlinear GSE**: We pioneer the application of Physics-Informed Neural Operators to solve the GSE with realistic, nonlinear GAQ current profiles. Unlike standard PINNs, our neural operator learning approach maps boundary shape parameters directly to equilibrium solutions, enabling rapid parameterized analysis across diverse plasma configurations without retraining.

2. **Comprehensive Architecture Benchmarking**: We conduct the first systematic comparison of state-of-the-art neural architectures within a unified neural operator framework. This benchmark identifies the Transformer-KAN Neural Operator (TKNO) as the superior architecture for capturing magnetic flux topologies.

3. **Rigorous Learning Paradigm Evaluation**: Through extensive out-of-distribution testing, we demonstrate that semi-supervised learning—integrating sparse labeled data with physics constraints—offers the optimal strategy. It resolves the "generalization crisis" of data-driven models, providing robust performance and "graceful degradation" in extrapolation regimes essential for reactor safety.

These findings establish a clear technical roadmap: semi-supervised PINO provides the accuracy, physical consistency, and robustness required for next-generation fusion reactor control systems.

The remainder of this paper is organized as follows: Section 2 details our research methodology, including overall framework (2.1), dataset generation (2.2), neural operator architectures (2.3), and training paradigms (2.4). Section 3 presents our experimental findings: supervised learning comparison (3.1-3.2), physics-informed methods (3.3), learning paradigm evaluation (3.4), and extrapolation analysis (3.5). Section 4 concludes with a summary of key findings, acknowledgment of limitations, and discussion of future research directions.

## 2 Physics-informed Neural Operator Learning Method

### 2.1 Overall Framework

The predictive modeling of complex physical systems, such as plasma equilibrium in tokamaks, is fundamentally rooted in solving PDE. Abstractly, a parameterized PDE can be expressed as the action of a differential operator, $\mathcal{N}$, on solution fields $u$, conditioned by a set of state parameters $f$:

$$\mathcal{N}(u(\mathbf{x}), f(\mathbf{x})) = 0, \quad \mathbf{x} \in \Omega \tag{1}$$

subject to appropriate boundary conditions, where $f \in \mathcal{F}$ and $u \in \mathcal{U}$ are from function spaces on a $d$-dimensional spatial domain $\Omega \in \mathbb{R}^d$. This formulation implicitly defines a solution operator $\mathcal{G}$: $\mathcal{F} \mapsto \mathcal{U}$, which maps the state parameters $f$ to the corresponding solution field $u$. The central challenge is to construct an efficient and accurate approximation $\mathcal{G}^\dagger$ of this potentially high-dimensional, nonlinear operator:

$$u = \mathcal{G}^\dagger\left(f\left(\mathbf{x}\right)\right) \tag{2}$$

Classical numerical solvers provide high-fidelity approximations by discretizing a continuous partial differential equation problem into an algebraic equation, but have substantial computational expense, arising from the need to solve large systems of equations for each new instance. Surrogate models based on standard deep learning architectures learn mappings between finite-dimensional Euclidean spaces ($\mathbb{R}^n \to \mathbb{R}^m$), effectively memorizing the relationship for a fixed discretization of the input and output. Consequently, they are inherently dependent on the specific mesh and fail to generalize to different resolutions or discretization. NO address this critical limitation by approximating the underlying solution operator $\mathcal{G}$ directly, which is designed to learn mappings between infinite-dimensional function spaces.

In our problem, the input parameters $f$ corresponds to the plasma shape parameters (denoted as $\lambda$), the solution $u$ corresponds to the poloidal magnetic flux (denoted as $\psi$), and the spatial coordinates is $\mathbf{x} = (R, Z)$. Therefore, the objective of neural operator learning is to construct a parameterized approximation to learn the mapping $\lambda \mapsto \psi(\mathbf{x})$ directly. To implement this, we propose a structured framework depicted in Figure 1, organized into three pivotal stages.

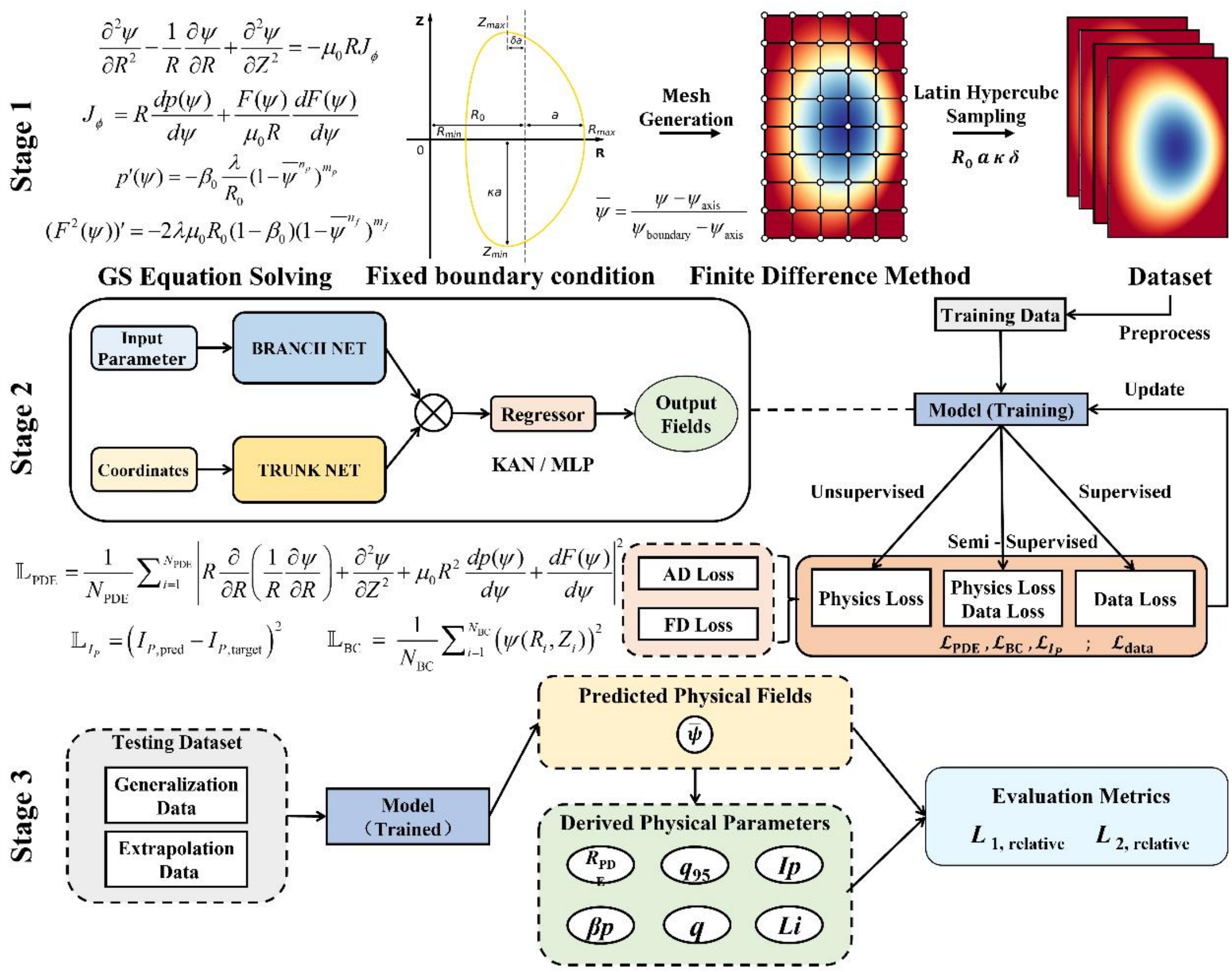

Figure 1. Overall framework

This initial stage is dedicated to building the high-quality dataset essential for training and validation. Because the true solution operator of the GSE lacks a known closed-form expression, we must approximate it empirically. To do this, we sample the operator's behavior by executing a high-fidelity numerical solver across a wide range of input parameters, thereby creating a comprehensive dataset of input-output examples. By creating $N$ distinct samples $\{f_i\}_{i=1}^{N}$ from the input parameter space, we generate a corresponding set of solution fields $\{u_i\}_{i=1}^{N}$. This process yields our dataset $\mathbb{D}$, which is a collection of input-output function pairs:

$$\mathbb{D} = \{(f_i, u_i)\}_{i=1}^{N} \text{ where } u_i = \mathcal{G}^{\dagger}(f_i) \tag{3}$$

The quality and breadth of this generated data form the essential bedrock upon which the NO is trained and tested.

Stage 2 represents the core of our research, where a predictive model is developed by training a sophisticated NO, $\mathbb{F}_{\Theta}$, to approximate the true solution operator $\mathcal{G}$, which is designed to map the input parameter function $f$ to the solution function $\hat{u}$:

$$\hat{u} = \mathbb{F}_{\Theta}(f) \tag{4}$$

The optimal learnable parameters $\Theta$ of the network are found by minimizing the expectation of a composite loss function over the data distribution:

$$\begin{cases} \Theta = \arg\min_{\Theta} \mathbb{E}_{(f,u)\sim\mathbb{D}}[\mathbb{L}_{\text{total}}(u, \mathbb{F}_{\Theta}(f); f)] \\ \mathbb{L}_{\text{total}}(u, \mathbb{F}_{\Theta}(f); f) = \omega_1 \mathbb{L}_{\text{data}}(u, \hat{u}) + \omega_2 \mathbb{L}_{\text{phy}}(\hat{u}; f) \end{cases} \tag{5}$$

where $(\lambda, \psi) \sim \mathbb{D}$ indicates samples obtained from the numerically generated dataset. The total loss function, $\mathbb{L}_{\text{total}}$ is composed of two key terms:

$\mathbb{L}_{\text{data}}$ is the numerical data loss, which quantifies the difference between the ground-truth solution $u$ and the network's prediction $\hat{u}$. It ensures the model's predictive accuracy. $\mathbb{L}_{\text{phy}}$ is the physics-informed loss, which measures the extent to which the predicted solution $\hat{u}$ satisfies the underlying GSE. It is formulated based on the PDE residual and other physical constrains, enhancing the model's generalization and physical fidelity.

By adjusting the weights $\omega_1$ and $\omega_2$, supervised ( $\omega_1 > 0,\ \omega_2 = 0$ ), unsupervised ( $\omega_1 = 0,\ \omega_2 > 0$ ), and semi-supervised ( $\omega_1 > 0,\ \omega_2 > 0$ ) training schemes are explored.

The final stage is designed to rigorously validate the trained operator predictive accuracy, generalization capability, and robustness. The model directly outputs the predicted physical field, $\hat{u} = \mathbb{F}_{\Theta}(f_{\text{new}})$. From this primary output, a set of other crucial physical parameters $\mathbb{P}_k$ are derived by applying specific post-processing operators or functionals, $\mathbb{H}_k$:

$$\mathbb{P}_k = \mathbb{H}_k(\hat{u}) \tag{6}$$

The model's performance is assessed on both interpolation data (within the training distribution) and more challenging extrapolation data (outside of it). To quantify its performance, standard relative error metrics are employed to compare the predicted fields and other physical parameters. This comprehensive evaluation is crucial for comparing the different training schemes and verifying the model's reliability and potential for real-world applications.

This comprehensive, multi-stage architecture provides a clear roadmap for developing fast, accurate, and physically consistent surrogate models for complex PDE systems. In the following, these three stages will be introduced in detail.

## 2.2 Physical model and Data Generation

### 2.2.1 Grad-Shafranov Equation

In standard cylindrical coordinates ($R$, $\phi$, $Z$), where axisymmetry implies no dependence on the toroidal angle $\phi$, the GSE defines the equilibrium via the poloidal magnetic flux function, $\psi$ ($R$, $Z$). The contours of the $\psi$ function represent the magnetic surfaces. The GSE is derived from the ideal-MHD force-balance equation, $\mathbf{J}\times\mathbf{B}=\nabla p$ . Full details of the derivation can be found in various plasma physics textbooks [38]. The standard form of the equation is [8]:

$$\Delta^*\psi=-\mu_0 R J_\phi \tag{7}$$

where $\mu_0$ is the vacuum permeability, $R$ is the radial coordinate and $J_\phi$ is the toroidal current density:

$$J_\phi = R\frac{dp(\psi)}{d\psi}+\frac{F(\psi)}{\mu_0 R}\frac{dF(\psi)}{d\psi} \tag{8}$$

where $p(\psi)$ is the plasma pressure, and $F(\psi) = RB_T$ is a function related to the poloidal current, where $B_T$ is the toroidal magnetic field. Both the pressure $p(\psi)$ and the function $F(\psi)$ are free functions that must be prescribed for a given equilibrium. The $\Delta^*$ operator is defined as:

$$\Delta^*\psi \equiv \frac{\partial^2\psi}{\partial R^2}-\frac{1}{R}\frac{\partial\psi}{\partial R}+\frac{\partial^2\psi}{\partial Z^2}=R\frac{\partial}{\partial R}\left(\frac{1}{R}\frac{\partial\psi}{\partial R}\right)+\frac{\partial^2\psi}{\partial Z^2} \tag{9}$$

### 2.2.2 Pressure and Poloidal Current Profiles

The source terms $\frac{dp(\psi)}{d\psi}$ and $F(\psi)\frac{dF(\psi)}{d\psi}$ define the toroidal plasma current density and thus determine the specific equilibrium configuration [39]. For the present study, these functions are parameterized using simple polynomial forms from GAQ code [40]. This approach is common in tokamak equilibrium analysis and reconstruction codes, including EFIT, which also originated from General Atomics. A normalized poloidal flux, $\overline{\psi}$, is defined to normalize the profiles:

$$\overline{\psi}=\frac{\psi-\psi_{\text{axis}}}{\psi_{\text{boundary}}-\psi_{\text{axis}}} \tag{10}$$

Here, $\psi_{\text{axis}}$ is the poloidal flux at the magnetic axis and $\psi_{\text{boundary}}$ is the flux at the plasma edge (typically set to 0). The source functions $p'(\psi)$ and $FF'(\psi)$ are then modeled as:

$$p'(\psi)=-\beta_0\frac{\lambda}{R_0}(1-\overline{\psi}^{n_p})^{m_p} \tag{11}$$

$$(F^2(\psi))'=-2\lambda\mu_0 R_0(1-\beta_0)(1-\overline{\psi}^{n_f})^{m_f} \tag{12}$$

Here, $R_0$ is the major radius of the device. The current profile is characterized by six free parameters: $n_p$, $m_p$, $n_f$, $m_f$, $\lambda$ and $\beta_0$. $n_p$, $m_p$, $n_f$ and $m_f$ control the shape of the pressure and current profiles. For the cases considered in this work, these exponents are held constant at $n_p = 2$, $m_p = 1$, $n_f = 1$ and $m_f = 1$. Although not explored in the present study, these parameters could also be treated as inputs for the neural network to learn in future work. The parameter $\lambda$ is a scaling factor related to the total plasma current $I_p$. The parameter $\beta_0$ is set to 0.8, a scalar quantity related to the poloidal beta, $\beta_p$. Compared to the Solov'ev equilibria commonly used in existing PINNs' studies [34], this polynomial model, while more challenging to solve due to its nonlinearity, offers a more realistic representation of actual tokamak plasmas, which has proven useful in both theoretical and experimental situations [40].

### 2.2.3 Boundary Condition and shape parameters

In this work, we consider a fixed-boundary scenario with an elongated, "D"-shaped toroidal geometry (Figure 2).

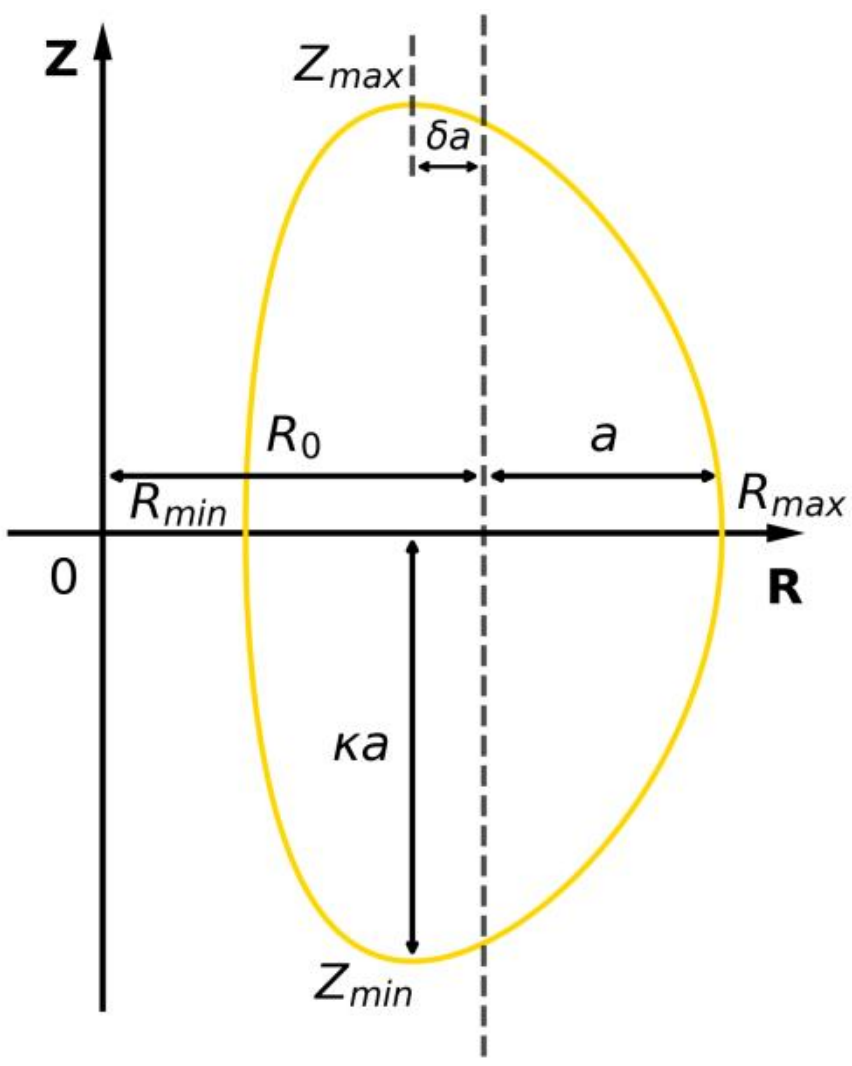

Figure 2. Elongated, "D"-shaped toroidal geometry

The plasma boundary ($\mathbf{\Gamma}$) can be parameterized by an angle $\tau \in [0, 2\pi]$ and four geometric shape parameters [41]:

$$R(\tau) = R_0 + a\cos(\tau + \arcsin(\delta)\sin(\tau)) \tag{13}$$

$$Z(\tau) = \varepsilon\kappa\sin(\tau) \tag{14}$$

These four parameters constitute our input state parameters $\boldsymbol{\lambda}$: major radius $R_0$, minor radius $a$, elongation $\kappa$, and triangularity $\delta$. The major radius $R_0$ and minor radius $a$ are given by:

$$R_0 = \frac{(R_{\max} + R_{\min})}{2} \tag{15}$$

$$a = \frac{\left(R_{\max} - R_{\min}\right)}{2} \tag{16}$$

These radii in turn define the aspect ratio $A = R_0 / a$. The plasma elongation $\kappa$ and $\delta$ triangularity are defined as:

$$\kappa = \frac{b}{a} \tag{17}$$

$$\delta = \frac{R_0 - R_{\mathrm{Zmax}}}{a} \tag{18}$$

For a fixed-boundary problem, the boundary condition is defined by setting the poloidal flux to zero on the plasma boundary $\mathbf{\Gamma}$:

$$\psi(R, Z)\,|_{\Gamma} = 0 \tag{19}$$

### 2.2.4 Numerical Solver

To solve the PDE, a numerical approach is employed based on a FDM coupled with a Successive Over-Relaxation (SOR) iterative scheme. The computational domain is a uniform rectangular grid in the ($R$, $Z$) plane and the grid boundaries are set at $R \in \left[R_0 - a, R_0 + a\right]$ and $Z \in \left[-\kappa a, \kappa a\right]$, ensuring that the D-shaped plasma boundary is fully contained within the computational domain for all combinations of shape parameters. The second-order partial differential GS operator $\Delta *$ in Eq. (9) is discretized on a 2D ($R$, $Z$) grid using a five-point central difference scheme. Specifically, the partial derivatives with respect to R and Z at each grid point $(R_i, Z_j)$ are approximated using second-order central differences. Combining these approximations yields the discretized form of the GSE:

$$\frac{\psi_{i+1,j} - 2\psi_{i,j} + \psi_{i-1,j}}{(\delta R)^2} + \frac{\psi_{i,j+1} - 2\psi_{i,j} + \psi_{i,j-1}}{(\delta Z)^2} - \frac{1}{R_i}\frac{\psi_{i+1,j} - \psi_{i-1,j}}{2\delta R} = S_{i,j} \tag{20}$$

where $\psi_{i,j} = \psi(R_i, Z_j)$, $\delta R$ and $\delta Z$ are the grid spacings, and the source term is given by:

$$S_{i,j} = -\mu_0 R_i^2 p'(\psi_{i,j}) - f(\psi_{i,j}) f'(\psi_{i,j}) \tag{21}$$

Where the pressure and poloidal current profile are computed by Eq. (11) and (12). This method transforms the continuous PDE into a large system of coupled non-linear algebraic equations for the values of psi at each grid point.

The resulting system of equations is solved iteratively using the Successive Over-Relaxation (SOR) method. For each grid point ($i$, $j$) within the plasma boundary, the value of the poloidal flux $\psi_{i,j}$ is updated at each iteration step $n$+1 based on the values from the previous step $n$ and the most recently updated neighboring points. The SOR update formula is given by:

$$\psi_{i,j}^{n+1} = (1-\omega)\psi_{i,j}^{n} + \frac{\omega}{C}\left( \frac{\psi_{i+1,j}^{n} + \psi_{i-1,j}^{n+1}}{(\delta R)^2} + \frac{\psi_{i,j+1}^{n} + \psi_{i,j-1}^{n+1}}{(\delta Z)^2} - \frac{\psi_{i+1,j}^{n} - \psi_{i-1,j}^{n+1}}{2R_i \delta R} - S_{i,j}^{n} \right) \tag{22}$$

where $C = \frac{2}{(\delta R)^2} + \frac{2}{(\delta Z)^2}$, and $\omega$ is the relaxation factor, which is chosen to optimize the convergence rate. The iteration continues until the relative error of the poloidal flux field between successive iterations falls below a tolerance or a maximum number of iterations is reached. The detailed introduction and code can be found in Xie's book [42].

### 2.2.5 Dataset Generation

To rigorously evaluate the PINO model, we generated two distinct datasets using the numerical solver described in Section 2.2.4. The input space consists of four plasma shape parameters $\lambda = (R_0, a, \kappa, \delta)$, sampled via Latin Hypercube Sampling [43] to ensure uniform coverage of the parameter space.

The two datasets are defined as follows:

1. Independent and Identically Distributed (IID) Dataset: Comprising 10,000 samples, this dataset represents standard tokamak operating regimes. It is partitioned into training (6,000), validation (2,000), and testing (2,000) subsets. The test set here evaluates the model's generalization capability within the training distribution.

2.Out-of-Distribution (OOD) Dataset: A separate set of 2,000 samples generated from significantly expanded parameter ranges (see Table 2). This dataset is strictly reserved for testing to assess the model's extrapolation robustness effectively simulating unseen physical scenarios.

Table 2. Parameter Ranges, Sizes, and Usage of the Datasets

| Dataset | Mode | Sample Size | $R_0$ | $a$ | $\kappa$ | $\delta$ |
|---|---|---|---|---|---|---|
| | Train | 6000 | | | | |
| IID | valid | 2000 | [1, 1.5] | [0.3, 0.5] | [1, 2.5] | [−0.5, 0.5] |
| | test | 2000 | | | | |
| OOD | test | 2000 | [0.5, 2.0] | [0.5, 2.0] | [0.5, 3.0] | [−0.9, 0.9] |

Each sample in the dataset is a tuple containing:

Input: The shape parameter vector $\lambda$ and the fixed spatial grid coordinates $(R, Z)$ of size 101 × 161 ($n_R = 101$, $n_Z = 161$). Output: The corresponding equilibrium magnetic flux field $\psi(R, Z)$.

Only numerical solutions that successfully converged (residual $< 10^{-8}$) were retained. These raw equilibrium solutions serve directly as the ground truth for supervised learning and validation, without additional preprocessing.

### 2.3 Model Architecture and Training Paradigms

#### 2.3.1 NO Learning Architecture

To construct an efficient surrogate model for the GSE, we propose a modular Neural Operator (NO) framework designed to learn the solution operator $\mathcal{G}$ defined in Section 2.1 that maps shape parameters $\lambda$ to the solution function $\psi(\mathbf{x})$. Inspired by the idea of DeepONet [30], our architecture decouples the learning of parameter dependencies and spatial dependencies into two parallel networks: a Branch Net and a Trunk Net. This design allows for flexible handling of the mapping between finite-dimensional parameter spaces and infinite-dimensional function spaces. As illustrated in Figure 3, the process consists of three stages:

1. Feature Encoding: The branch net $\mathcal{B}$ encodes the parameters $\lambda$ into a latent feature vector $\mathbf{b} \in \mathbb{R}^{p}$, where $p$ is the dimension of latent space. Simultaneously, the trunk net $\mathcal{T}$ maps the spatial coordinate grid $\mathbf{x}$ (for $d$= 2, $\mathbf{x} \in \mathbb{R}^{2}$) into a high-dimensional basis function space $\mathbf{T}$ (for $d$= 2, $\mathbf{T} \in \mathbb{R}^{H \times W \times p}$).

2. Feature Fusion: We follow the standard DeepONet formulation, which combines these two representations via a dot product.

3. Regression: A Regressor $\mathcal{R}$ projects the fused features to the final scalar field $\psi$. The overall inference process can be mathematically formalized as:

$$\mathcal{G}^{\dagger}(\boldsymbol{\lambda})(\mathrm{x}) \approx \mathcal{R}\left(\mathcal{B}(\lambda) \otimes \mathcal{T}(\mathbf{x})\right) \tag{23}$$

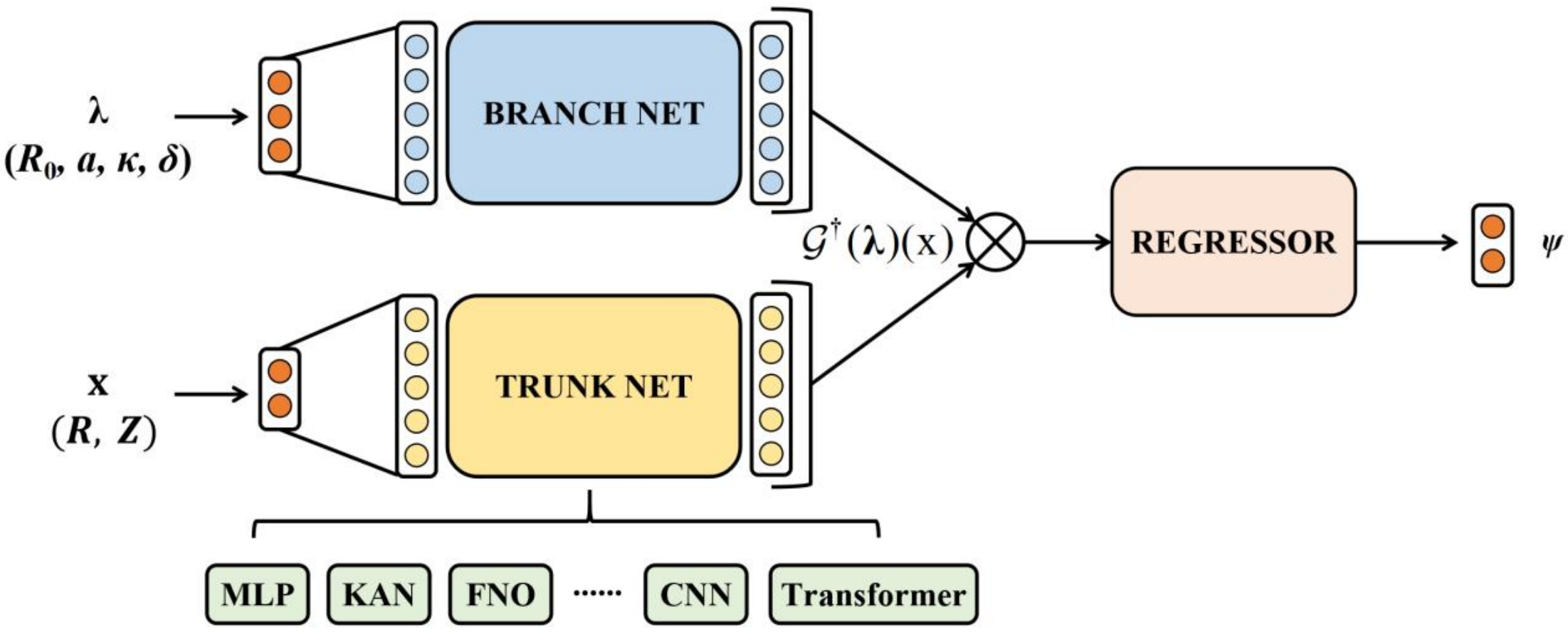

Figure 3. NO architecture.

This modular structure stands in contrast to a standard neural network architecture, where all inputs—both source parameters and spatial coordinates—are typically concatenated and passed through a single, monolithic network. The key advantage of our framework is its flexibility, each of the three modules can be implemented using a different network architecture. This allows us to select the most appropriate architecture to capture the distinct characteristics of the data and relationships at each stage.

For the specific task of solving the GSE, we apply this framework as follows in which the inputs are plasma shape parameters and coordinates, and the outputs are the magnetic flux fields. The input parameters are a 4-dimensional vector of plasma shape parameters $\boldsymbol{\lambda} = [R_0, a, \kappa, \delta]$. Since this input space is low-dimensional and the mapping is relatively simple, we employ a standard MLP for the branch net. In contrast, the trunk net must process the 2D spatial coordinates ($R$, $Z$) and capture the complex, non-linear spatial relationships of the magnetic flux field. Given this critical role, we employ this module as the testbed for benchmarking five distinct deep learning architectures. The selection of these architectures is not arbitrary but aims to explore which types of prior knowledge (e.g., spatial multi-scale structures, spectral sparsity or global relationship dependencies) are most effective for describing the tokamak equilibrium state.

While the regressor component is often a simple feed-forward network, to enhance the model's ability to represent the final solution field, in this paper, we compare two distinct structures including MLP and KAN.

To clearly present our results in figures and tables, we introduce a concise {trunk net}-{regressor} naming convention based on the following abbreviations: M (MLP), K (KAN), F (FNO), C (CNN), T (Transformer). Thus, a NO with a Transformer Trunk Net and a KAN Regressor is abbreviated as T-K. This standardized naming will be used in all subsequent figures and tables.

In the following sections, we will provide a detailed introduction to the architectures evaluated for the Trunk Net.

### 2.3.2 KAN

The KAN is a novel neural network architecture inspired by the Kolmogorov-Arnold representation theorem [44], proposed as a potential alternative to Multi-Layer Perceptron (MLP). A typical MLP computes the $l$-th layer output by:

$$\mathbf{x}_l = \sigma(\mathbf{W}_l \mathbf{x}_{l-1} + \mathbf{b}_l) \tag{24}$$

where $\mathbf{W}_l$ and $\mathbf{b}_l$ are the trainable weight matrix and bias of the $l$-th layer; $\sigma$ is a fixed nonlinear activation function.

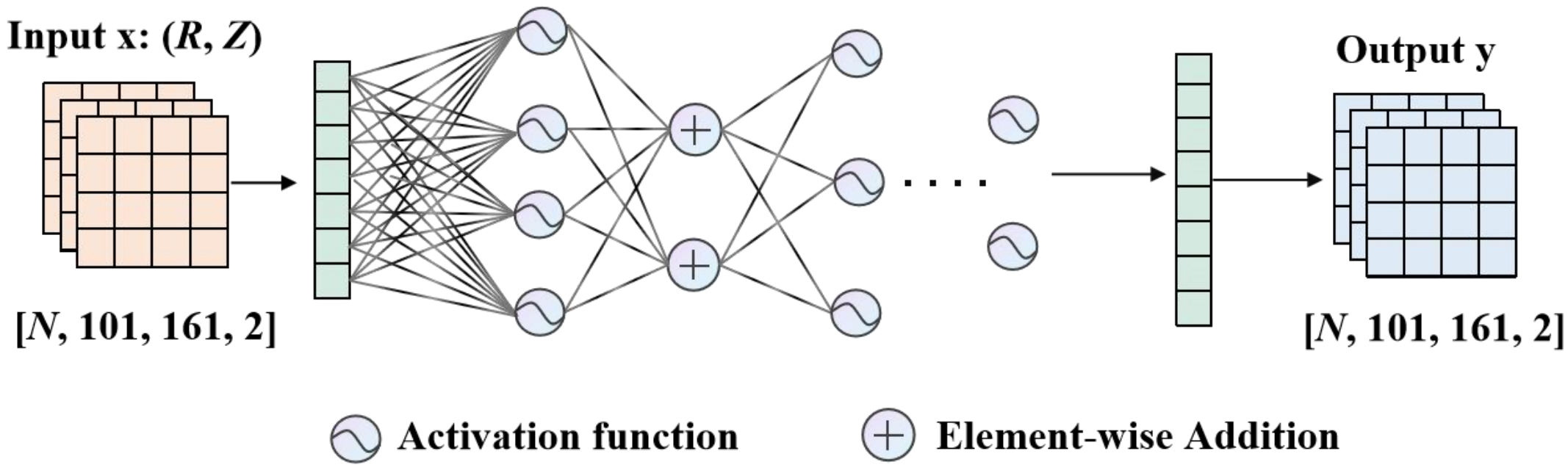

Figure 4. KAN structure.

In contrast, KAN (Figure 4) replaces these fixed activations with learnable univariate functions on the edges (weights). For an input $\mathbf{x} = (x_1, \ldots, x_n)$ mapped to an output $\mathbf{y} = (y_1, \ldots, y_m)$, the $j$-th output neuron $y_j$ is calculated as follows:

$$\mathbf{y}_j = \sum_{i=1}^{n} \phi_{j,i}\left(x_i\right) = \sum_{i=1}^{n} (w_b \cdot b\left(x_i\right) + w_s \cdot \sum_{k=1}^{G+K-1} c_k B_k\left(x\right)) \tag{25}$$

where $\phi_{j,i}$ is the learnable function connecting the $i$-th input neuron and the $j$-th output neuron; $b(x)$ is a basis function (e.g., SiLU); $c_k$ are the learnable spline coefficients with $G$ grid intervals and spline order $K$. B-splines provide smooth and flexible local adjustment, and stacking such layers forms the KAN, which offers stronger approximation ability and better interpretability than MLPs.

### 2.3.3 FNO

The Fourier Neural Operator (FNO) offers a distinct paradigm by parameterizing the integral operator in the frequency domain. As illustrated in Figure 5, the FNO model $\mathcal{T}_{\text{FNO}}$ is formulated as a composition of three operators: a lifting operator $\mathcal{P}$ which maps, a stack of $L$ Fourier layers $\mathcal{F}^L$, and a projection operator $\mathcal{Q}$:

$$\mathbf{T} = \mathcal{T}_{\text{FNO}}(\mathbf{x}) = \mathcal{Q}\left(\mathcal{F}^L(\mathcal{P}(\mathbf{x}))\right); \quad \mathcal{F}^L = \mathcal{F}_L \circ \ldots \circ \mathcal{F}_1 \tag{26}$$

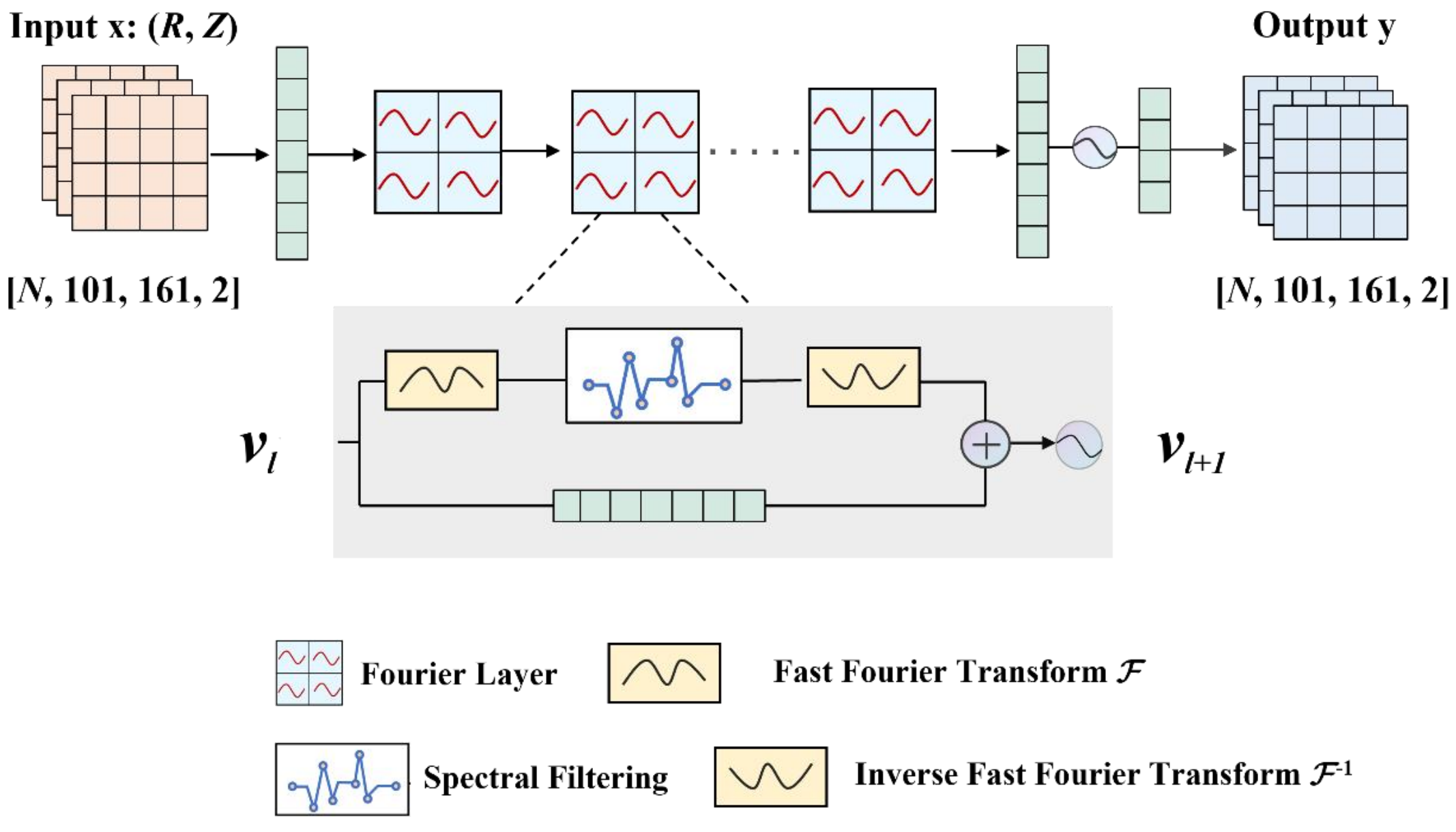

Figure 5. FNO structure.

Unlike CNNs, FNO is resolution-invariant and possesses a global receptive field [45] .The fundamental component is the Fourier Layer, which transforms the input to the frequency domain via FFT, applies a learnable operator $R_\phi$ on low-frequency modes, and brings it back with IFFT. The result is added to a local mapping $Wv_l(x)$ and passed through an activation $\sigma$, yielding the update:

$$v_{l+1}(x) = \sigma(Wv_l(x) + \mathcal{F}^{-1}(R_\phi \cdot (\mathcal{F}v_l))(x)) \tag{27}$$

The lifting operator $\mathcal{P}$ maps the input function $\mathbf{x}$ from the physical parameter space to a higher-dimensional latent feature space, producing the initial feature field $v_0(x) = \mathcal{P}(\mathbf{x})$ , and the lifted field $v_0(x)$ is then iteratively updated through subsequent Fourier layers, yielding intermediate representations $v_l(x)$.

### 2.3.4 CNN

CNNs are a class of deep learning models renowned for their exceptional performance on grid-structured data, such as images. The core principle of CNNs is to extract local spatial features through convolutional kernels and to build a hierarchy of features—from low-level to high-level—by stacking convolutional and pooling layers.

In scientific computing, particularly for solving PDE, CNNs are widely used to learn the complex spatial distributions of physical fields. For the Trunk Net, we adopt a U-Net architecture (Figure 6), which features a symmetric encoder-decoder structure with skip connections. This design allows the network to learn multiscale representations: the encoder captures global context through down-sampling, while the decoder recovers high frequency details (gradients) essential for the magnetic flux function via up-sampling.

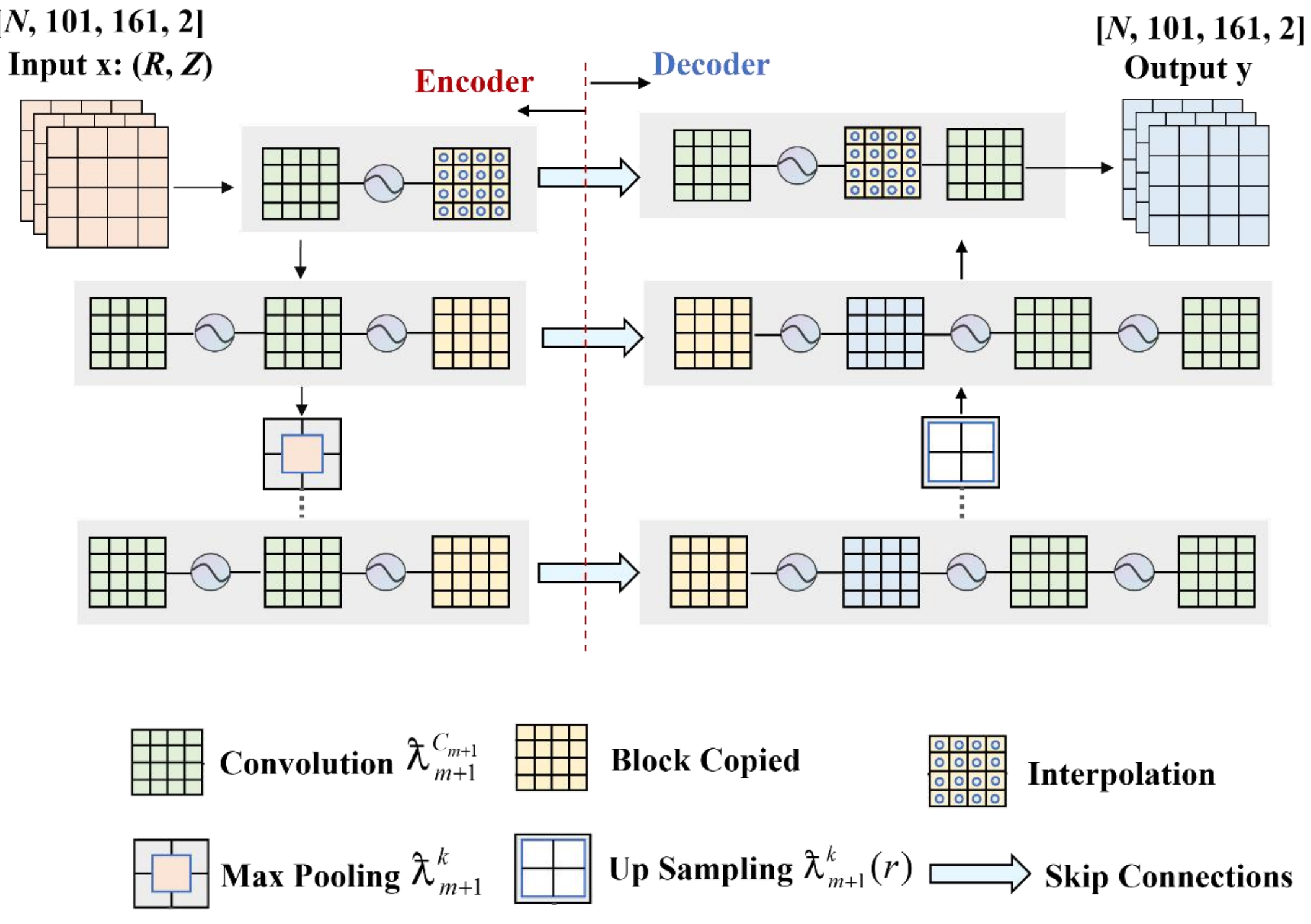

Figure 6. CNN (U-Net) structure.

The overall U-Net mapping $\mathcal{T}_{\text{U-Net}}$ can be abstractly represented as a nested composition of encoder ( $\mathcal{E}$ ) and decoder () blocks:

$$\mathbf{T} = \mathcal{T}_{\text{U-Net}}(\mathbf{X}) = \mathcal{D}_1\left(\mathcal{E}_1(\mathbf{X}), \mathcal{D}_2(\mathcal{E}_2(\ldots))\right) \tag{28}$$

where the skip connections pass feature maps from encoder block $\mathcal{E}_i$ to the corresponding decoder block $\mathcal{D}_i$ . For a feature map $\mathbf{h}^{(l-1)}$ at layer $l-1$, the output $\mathbf{h}^{l}$ is computed via learnable kernels **W** and bias **b**:

$$\mathbf{h}^{(l)} = \sigma\left(\mathbf{W}^{(l)} * \mathbf{h}^{(l-1)} + \mathbf{b}^{(l)}\right) \tag{29}$$

### 2.3.5 Transformer

The Transformer architecture, originally developed for natural language processing, relies entirely on a self-attention mechanism instead of recurrence or convolution to efficiently model long-range dependencies [46].

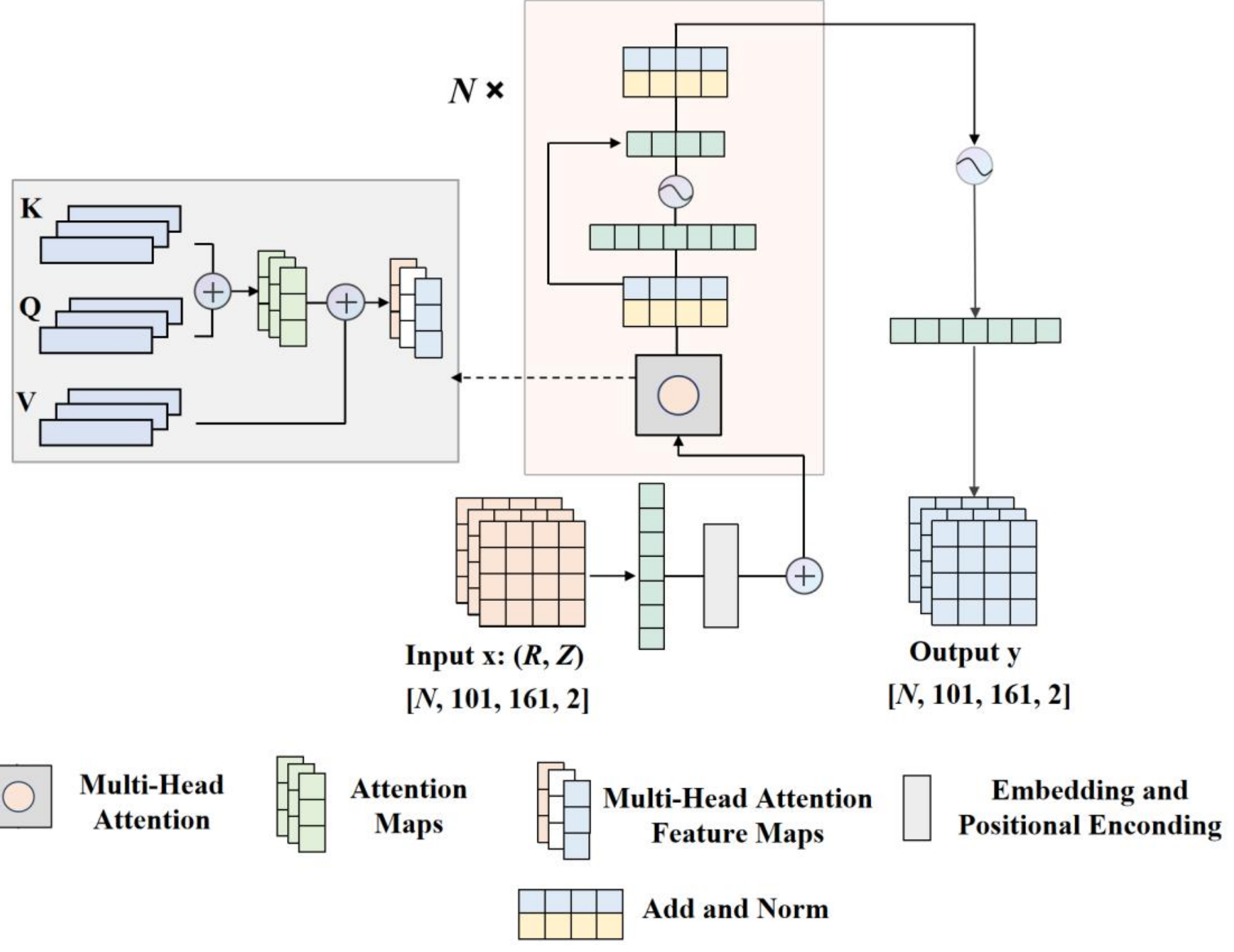

Figure 7. Transformer structure.

As shown in Figure 7 the overall mapping $\mathcal{T}_{\text{Trans}}$ consists of an Embedding layer $\mathcal{E}$, a stack of Transformer Encoders $\mathbf{T}_{\text{Enc}}$, and a final mapping layer $\mathcal{M}$:

$$\mathbf{T} = \mathcal{T}_{\text{Trans}}(\mathbf{X}) = \mathcal{M}\left(\mathbf{T}_{\text{Enc}}(\mathcal{E}(\mathbf{X}))\right) \tag{30}$$

In this framework, the spatial coordinate grid **X** is treated as a sequence of tokens. The key innovation is the Scaled Dot-Product Attention mechanism within each encoder, which computes a weighted sum of values (**V**) based on the compatibility of queries (**Q**) and keys (**K**):

$$\text{Attention}(\mathbf{Q}, \mathbf{K}, \mathbf{V}) = \text{softmax}\left(\frac{\mathbf{Q}\mathbf{K}^T}{\sqrt{d_k}}\right)\mathbf{V} \tag{31}$$

where the **Q** (Query), **K** (Key), and **V** (Value) are obtained from linear transformations of the input embeddings, and $d_k$ is the dimension of **K**. Multi-head attention computes this process in parallel across multiple heads and concatenates the outputs.

To further enhance the efficiency and scalability of neural operator, we also adopt the Galerkin-type attention mechanism [47], which is computed as follows:

$$\left(\mathbf{z}^m\right)_n = \sum_{i=1}^{h} \frac{\left(\mathbf{k}^i \cdot \mathbf{v}^m\right)}{w} \left(\mathbf{q}^i\right)_n \approx \sum_{i=1}^{h} \left(\int_{\Omega} \left(k_i(\zeta) v_m(\zeta)\right) d\zeta\right) q_i\left(x_n\right) \tag{32}$$

where $m$ and $n$ denote the $m$-th column vector of the corresponding matrix and the $n$-th element of that column vector, respectively. $\zeta$ represents the feature maps of $\mathbf{Q}$.

### 2.3.6 Training Paradigms

As illustrated in Eq. (5) , the predictive model is trained by minimizing a composite loss function, which is constructed from two primary components: a Data Loss and a Physics Loss. By selectively combining these components, we can explore three distinct training paradigms: supervised (data-driven), unsupervised (physics-informed), and semi-supervised. This section provides a detailed introduction to the loss function formulations. A key focus will be on the implementation of the PINO framework, particularly clarifying the differences between using Automatic Differentiation (AD) and the Finite Difference (FD) to enforce physical constraints.

In supervised learning, model training relies exclusively on the data pairs generated by the numerical solver. The loss function comprises only a data-fitting term, typically the Mean Squared Error (MSE), which quantifies the discrepancy between the network's prediction and the ground-truth solution:

$$\mathbb{L}_{\text{data}} = \frac{1}{N}\sum_{i=1}^{N}\left(\frac{1}{M}\sum_{j=1}^{M}|\psi_{\text{pred},i,j} - \psi_{\text{true},i,j}|^2\right) \tag{33}$$

where $N$ is the number of points in the training samples and $M$ is the number of mesh points.

In contrast, PINO (Figure 8) is a semi-supervised or unsupervised learning method that directly integrates physical laws describing the PDE residual as a regularization term into the neural network training process. The core advantage of this method is its ability to leverage unlabeled sampling points within the physical domain to constrain the model's behavior, thereby guiding the network to learn physically self-consistent solutions and improving the model’s generalization ability and physical fidelity, especially in cases of sparse or missing data [25].

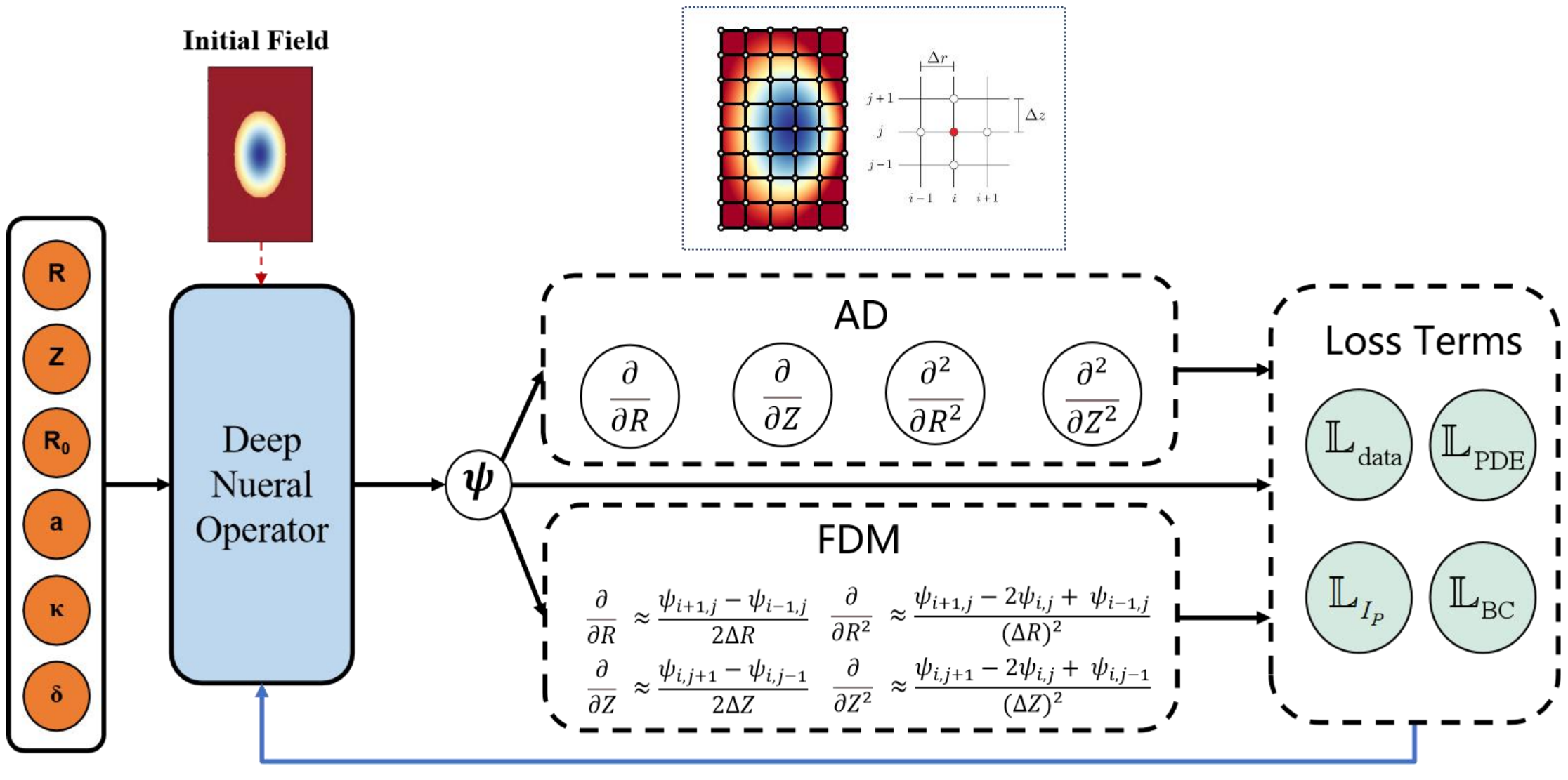

Figure 8. PINO architecture.

In this study, the total loss function is a weighted sum of several components:

$$\mathbb{L}_{\text{total}} = \omega_{\text{data}}\mathbb{L}_{\text{data}} + \left(1-\omega_{\text{data}}\right)\mathbb{L}_{\text{phs}} \tag{34}$$

$$\mathbb{L}_{\text{phs}} = \omega_{\text{PDE}}\mathbb{L}_{\text{PDE}} + \omega_{I_p}\mathbb{L}_{I_p} + \omega_{\text{BC}}\mathbb{L}_{\text{BC}} \tag{35}$$

where $\omega$ are the weight coefficients for each term, used to balance the contribution of different loss terms. The terms are defined as follows:

$$\mathbb{L}_{\text{PDE}} = \frac{1}{N_{\text{PDE}}}\sum_{i=1}^{N_{\text{PDE}}}\left|R\frac{\partial}{\partial R}\left(\frac{1}{R}\frac{\partial\psi}{\partial R}\right)+\frac{\partial^2\psi}{\partial Z^2}+\mu_0 R^2\frac{dp(\psi)}{d\psi}+\frac{dF(\psi)}{d\psi}\right|^2 \tag{36}$$

$$\mathbb{L}_{I_p} = \left(I_{p,\text{pred}} - I_{p,\text{target}}\right)^2,\ I_p = \iint J_\phi dRdZ \tag{37}$$

$$\mathbb{L}_{\text{BC}} = \frac{1}{N_{\text{BC}}}\sum_{i=1}^{N_{\text{BC}}}\left(\psi(R_i,Z_i)\right)^2 \tag{38}$$

The primary component, $\mathbb{L}_{\text{PDE}}$, calculates the mean squared residual of the GSE over $N_{\text{PDE}}$ internal collocation points. Computing this requires first and second-order derivatives of $\psi_{\text{pred},i}$, which are either computed via AD for architectures like MLPs and KANs, or approximated using FD for models like FNO and Transformers where AD is problematic. The $p'(\psi)$ and $FF'(\psi)$ terms are calculated by the current profile model in Eq. (5) and Eq. (6). To ensure global physical accuracy, the $\mathbb{L}_{I_p}$ term penalizes the difference between the predicted current ($I_p$, the integral of the toroidal current density $J_\phi$ over the plasma cross-section) and its target current (fixed at $3\times10^6$ A in this work) after normalized. For fixed-boundary problem, $\mathbb{L}_{\text{BC}}$ enforces the boundary conditions, typically by applying a Dirichlet boundary condition (such as $\psi$

= 0) at the outermost magnetic surface of the plasma for fixed-boundary problem.

Due to the presence of source terms in the GSE, initializing the network to output a field resembling a Gaussian distribution can sometimes aid in the initial stages of PINO training, preventing local optimization.

To ensure the reproducibility of our results, detailed hyperparameters for the network architecture and training process are listed in Table 3. All models were implemented in PyTorch and trained on a single NVIDIA RTX 5090 GPU. Due to the optimal performance, we present the TKNO architecture parameters and other model structures have similar numbers of total parameters. Training was performed using the Adam optimizer. We utilized a fixed weighting strategy for the loss function components, where the PDE residual weight was tuned to $1\times10^{-5}$ to balance physical consistency with data fidelity. The learning rate followed a multi-step decay schedule to ensure precise optimization in the final stages.

Table 3. Detailed Hyperparameter Settings for the TKNO Training

| Category | Parameter | Value |
|---|---|---|
| **Optimization** | Optimizer | Adam ($\beta_1$ = 0.8, $\beta_2$ = 0.9) |
| | Learning Rate (Initial) | $1\times10^{-3}$ |
| | LR Scheduler | MultiStepLR ($\gamma$ = 0.1) (milestone = 150 for supervised, [500, 1000, 1500] for others) |
| | Random Seed | 2025 |
| **Loss Weights** | Strategy | fixed |
| | $\omega_{\text{data}}$ | 1.0 (0.5 for semi-supervised) |
| | $\omega_{\text{PDE}}$ | $1\times10^{-5}$ |
| | $\omega_{\text{BC}}, \omega_{I_p}$ | 1.0 |
| **Architecture** | Branch Net (MLP) | layers: 3; width: 64 |
| | Trunk Net (Transformer) | encoder layers: 3; width: 64 |
| | Regressor (MLP/KAN) | layers: 1; width: 16 |

## 2.4 Inference Results and Evaluation Metrics

To comprehensively and deeply evaluate the performance of different deep learning models in solving the GSE, we have designed a hierarchical evaluation system. This system not only includes an assessment of the predictive accuracy of the model's direct output, the magnetic flux function $\psi$, it also includes an evaluation of a series of key macroscopic physical quantities derived from $\psi$ and its derivatives. These metrics therefore provide a more profound insight into whether the model has truly captured the intrinsic physical laws of tokamak equilibrium.

### 2.4.1 Magnetic Flux Field Prediction Accuracy

This is the most direct assessment of a model's performance, measuring the consistency between the predicted two-dimensional magnetic flux field $\psi_{\text{pred}}$ and the true magnetic flux field $\psi_{\text{true}}$ calculated by a high-precision numerical solver. We use two standard mean relative error metrics:

$$\mathcal{L}_{1,\text{rel}}(\psi) = \frac{1}{N}\sum_{i=1}^{N}\frac{\left|\psi_{\text{pred}} - \psi_{\text{true}}\right|}{\left|\psi_{\text{true}}\right|} \tag{39}$$

$$\mathcal{L}_{2,\text{rel}}(\psi) = \frac{1}{N}\sum_{i=1}^{N}\frac{\left|\psi_{\text{pred}} - \psi_{\text{true}}\right|^2}{\left|\psi_{\text{true}}\right|^2} \tag{40}$$

In the above formulas, the summation is over all $N$ points on the evaluation grid.

### 2.4.2 PDE Residual Error

The PDE residual error is the $L_1$ and $L_2$ norm of the GSE residual calculated by FD over the predicted flux field:

$$\mathcal{L}_{1,\text{ PDE}} = \frac{1}{N_{\text{PDE}}}\sum_{i=1}^{N_{\text{PDE}}}\left|R\frac{\partial}{\partial R}\left(\frac{1}{R}\frac{\partial\psi}{\partial R}\right) + \frac{\partial^2\psi}{\partial Z^2} - R\frac{dp(\psi)}{d\psi} + \frac{F(\psi)}{\mu_0 R}\frac{dF(\psi)}{d\psi}\right| \tag{41}$$

$$\mathcal{L}_{2,\text{ PDE}} = \frac{1}{N_{\text{PDE}}}\sum_{i=1}^{N_{\text{PDE}}}\left|R\frac{\partial}{\partial R}\left(\frac{1}{R}\frac{\partial\psi}{\partial R}\right) + \frac{\partial^2\psi}{\partial Z^2} - R\frac{dp(\psi)}{d\psi} + \frac{F(\psi)}{\mu_0 R}\frac{dF(\psi)}{d\psi}\right|^2 \tag{42}$$

### 2.4.3 Derived Physical Parameters

While the magnetic flux function $\psi$ (R, Z) provides the fundamental description of the magnetic topology, the equilibrium state is further characterized by a set of macroscopic physical quantities derived from $\psi$ and its spatial derivatives. These parameters are critical for assessing the plasma's operational boundaries, confinement quality, and magnetohydrodynamic (MHD) stability. Consequently, verifying the model's accuracy in predicting these derived quantities is as important as the flux field itself.

**Total Plasma Current ($I_p$)**

The total plasma current $I_p$ serves as a fundamental global constraint for the equilibrium. It is obtained by integrating the toroidal current density over the entire poloidal cross-sectional area of the plasma in Eq. (37). Accurate prediction of $I_p$ ensures that the model correctly captures the total magnetic energy and the overall magnitude of the poloidal field.

**Safety Factor ($q$) and $q_{95}$**

The safety factor profile, $q(\psi)$, describes the helicity of the magnetic field lines, defined as the limit of the number of toroidal turns a field line makes per poloidal turn. It is a dimensionless measure of the magnetic field line pitch and is topologically invariant:

$$q(\psi) = \frac{f(\psi)}{2\pi} \oint_{\psi} \frac{dl}{RB_p} \tag{43}$$

where $f(\psi) = RB_{\psi}$ is the poloidal current function and the line integral is taken along a contour of constant poloidal flux. Of particular importance for stability analysis is the parameter $q_{95}$, defined as the safety factor evaluated at the magnetic flux surface enclosing 95% of the normalized poloidal flux ($\psi_{95}$):

$$q_{95} \equiv q(\psi)\big|_{\psi=\psi_{95}} \tag{44}$$

In diverted tokamak configurations, the safety factor diverges to infinity at the separatrix ($q_{\text{sep}} \to \infty$). Therefore, $q_{95}$ serves as a robust representative metric for the edge magnetic topology and is strictly correlated with the stability limits of external kink modes and the onset of disruptions.

**Poloidal Beta ($\beta_p$)**

The poloidal beta, $\beta_p$ measures the efficiency of the poloidal magnetic field in confining the plasma pressure. It is defined as the ratio of the volume-averaged plasma pressure to the magnetic pressure associated with the poloidal magnetic field at the boundary:

$$\beta_p = \frac{2\mu_0 \langle p \rangle_V}{\langle B_p \rangle_a^2} \tag{45}$$

where $\langle p \rangle_V$ is the volume-averaged plasma pressure, and $\langle B_p \rangle_a$ is the poloidal magnetic field averaged over the plasma boundary.

**Normalized Internal Inductance ($l_i$)**

The normalized internal inductance, $l_i$, quantifies the magnetic energy stored within the plasma volume relative to the poloidal field energy at the surface. It effectively describes the "peakedness" of the toroidal current density profile:

$$l_i = \frac{\langle B_p^2 \rangle_V}{\langle B_p \rangle_a^2} \tag{46}$$

where $\left\langle B_p^2 \right\rangle_V$ is the volume average of the square of the poloidal magnetic field. Physically, a lower $l_i$ corresponds to a broader current distribution (typical of "skin" currents), while a higher $l_i$ indicates a more centrally peaked current profile. This parameter plays a vital role in determining the vertical stability of elongated plasmas.

## 3 Results and Discussions

This section presents a comprehensive analysis of the performance of various neural network architectures for solving the Grad-Shafranov (GS) equation. We first compare the predictive accuracy of mainstream neural network structures in the supervised learning paradigm.

We then develop and evaluate a PINO using a FD method to ensure physical consistency. Meanwhile, we investigate a semi-supervised approach that combines the strengths of both supervised and unsupervised models.

Finally, we compared the generalization ability and extrapolation ability of three learning paradigms. The first experiment compared the test results under different training set sizes. The second experiment tested the reasoning results of the models outside the training range.

### 3.1 Performance Comparison of Neural Network Structures

To identify the optimal surrogate model architecture, a comprehensive ablation study was designed to evaluate the performance impact of different network components, specifically comparing: (1) general neural networks vs. NO; (2) different trunk net structures (MLP, KAN, FNO, CNN, Transformer); and (3) different regressor structures (MLP vs. KAN). We evaluated the models first by their training dynamics (Figure 9) and then by their generalization performance on the test dataset (Figure 10 and Table 4).

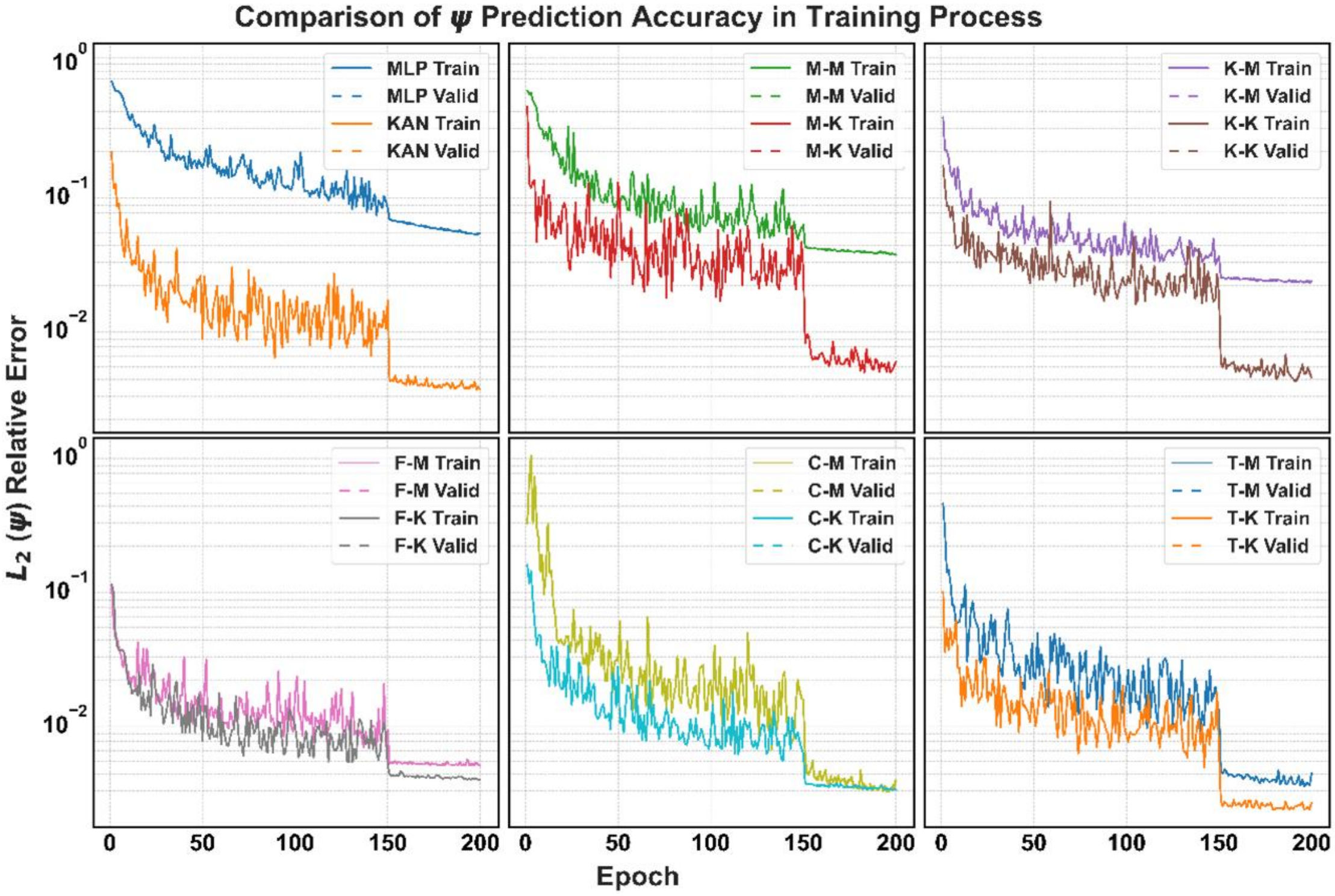

Figure 9. Comparison of the L2 relative error for the magnetic flux field ψ on the train and valid datasets, testing standard networks and various NO {trunk net}-{regressor} combinations.

The training curves in Figure 9 show that all models train stably, with consistent performance between their training and validation sets. The standard MLP (top-left, blue) converges slowly to the highest error, while the KAN baseline (top-left, orange) performs significantly better. The NO models using advanced trunk nets—FNO, CNN, and Transformer (bottom row)—achieve superior performance. Regarding the regressor, KAN also outperforms MLP; however, this improvement becomes marginal when combined with advanced trunk nets. The sharp error drop at epoch 150 results from scheduled learning rate decay.

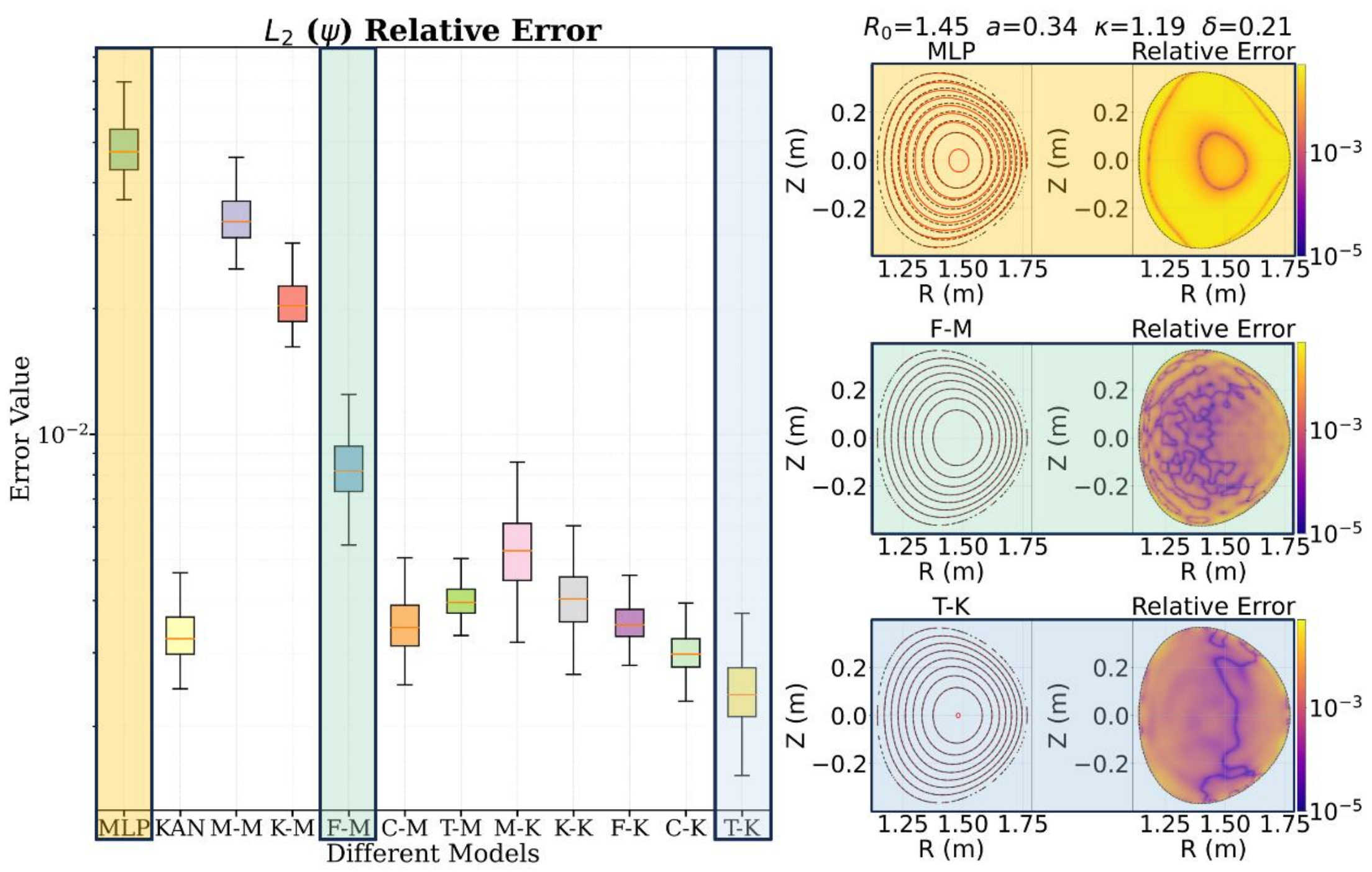

Figure 10. Prediction performance comparison of magnetic flux ($\psi$) on the test set. (Left) Box plot of the $L_2$ relative error in magnetic flux ($\psi$) for different models. (Right) Qualitative case studies for MLP, F-M, and T-K, showing predicted flux contours (red dashed) vs. ground truth (black solid) (left pane) and the relative error map (right pane).

The test set results in Figure 10 and Table 4 reinforce these findings. The box plot (Figure 10) quantitatively confirms that the MLP performs worst. Neural operators with advanced trunk nets (FNO, CNN, Transformer) generally outperform the simpler MLP and KAN-based trunks, indicating their superior ability to capture complex spatial non-linearities. To our surprise, the standard KAN model performed better than all neural operators with MLP regressor. Therefore, a single layer of KAN is a better choice for regressor and T-K model achieves the lowest median $L_2$ relative error. The qualitative case studies (Figure 10, right) visually support this: the MLP's prediction deviates massively, showing large, diffuse areas of high error. In contrast, both the F-M and T-K models' predicted flux contours (red dashed) are nearly indistinguishable from the ground truth (black solid). However, a closer look at their relative error maps reveals that the F-M model exhibits slightly higher error, with scattered high-error spots in the core and more pronounced error near the plasma boundary, while the T-K model maintains a smoother, lower error field. This performance stems from a structural synergy: the Transformer's self-attention mechanism effectively captures the global long-range dependencies inherent in the magnetic topology. Complementing this, the KAN regressor's learnable splines provide superior capacity for resolving the strong local non-linearities in the flux distribution compared to fixed activation functions.

Table 4. Performance comparison of different supervised models

| Model | Parameters (M) | Train Dataset | | Test Dataset | | |
|---|---|---|---|---|---|---|
| | | $L_1$ ($\psi$) | $L_2$ ($\psi$) | $L_1$ ($\psi$) | $L_2$ ($\psi$) | $L_1$(PDE) |
| MLP | **0.13** | 5.14% | 4.92% | 5.15% | 4.93% | 4.982 |
| KAN | 0.34 | 0.30% | 0.33% | 0.30% | 0.34% | 7.860 |
| M-M | 0.20 | 3.38% | 3.41% | 3.42% | 3.45% | 5.642 |
| K-M | 0.47 | 2.06% | 2.15% | 2.10% | 2.19% | 6.879 |
| F-M | 0.30 | 0.91% | 0.87% | 0.93% | 0.88% | 10.103 |
| C-M | 0.44 | 0.35% | 0.36% | 0.36% | 0.36% | 31.010 |
| T-M | 0.28 | 0.41% | 0.40% | 0.41% | 0.40% | 8.018 |
| M-K | 0.24 | 0.56% | 0.54% | 0.56% | 0.54% | 7.968 |
| K-K | 0.49 | 0.39% | 0.41% | 0.30% | 0.42% | 7.873 |
| F-K | 0.30 | 0.31% | 0.37% | 0.31% | 0.37% | 9.796 |
| C-K | 0.47 | 0.26% | 0.28% | 0.26% | 0.29% | 17.293 |
| **T-K** | 0.29 | **0.23%** | **0.25%** | **0.23%** | **0.25%** | **3.556** |

Table 4 provides the final quantitative metrics. All models are lightweight, with parameters in the 0.1-0.5 million range. The T-K model proves to be the most effective, achieving the lowest test $L_2$ ($\psi$) relative error (0.25%) with a moderate parameter count (0.29 M), with the C-K model being a close second (0.29% relative error).

However, all models achieving poor performance on PDE residual, far greater than the negligible error ($<10^{-4}$) of the numerical solver. This discrepancy highlights a common challenge in surrogate modeling for PDEs: a model can excel at fitting data points (low data loss) without accurately learning the underlying differential structure or physical laws governing the system. While neural networks are powerful function approximators, accurately learning the second-order derivatives involved in the GSE from data alone is substantially more difficult than learning the function values themselves, especially when the training data may contain numerical noise from the original solver. This gap between data fidelity and physical consistency provides a strong motivation for exploring PINO.

Considering the data fidelity, the T-K neural operator (TKNO) is identified as the optimal supervised surrogate in this evaluation. Therefore, in the following section, we will adopt this T-K architecture as the foundation for exploring the performance of a PINO.

### 3.2 Development and Evaluation of Unsupervised PINO

To enforce physical consistency, an unsupervised PINO was constructed based on the high-performing TKNO architecture. As explained in 2.3.6, Transformers are not readily compatible with the AD, a FD Method was utilized to compute the derivatives for the PDE loss. The total loss function is a composite of three terms including $\mathcal{L}_{\text{PDE}}$, $\mathcal{L}_{I_p}$ and $\mathcal{L}_{\text{BC}}$ shown in Eq. (34).

Balancing these terms is crucial for successful training, as their gradients can point in conflicting directions, especially given the non-linear nature of the source term in the GSE. The loss weights are firstly tuned on a single case before extending the training to the full parametric space of 6000 samples. As shown in Figure 11, the weight of the PDE loss term significantly impacts the outcome. A large PDE weight can cause the training to fall into a local minimum that deviates from the true solution, while a smaller PDE weight allows for a balanced optimization. A weight of $1\times10^{-5}$ for the PDE loss was found to be optimal.

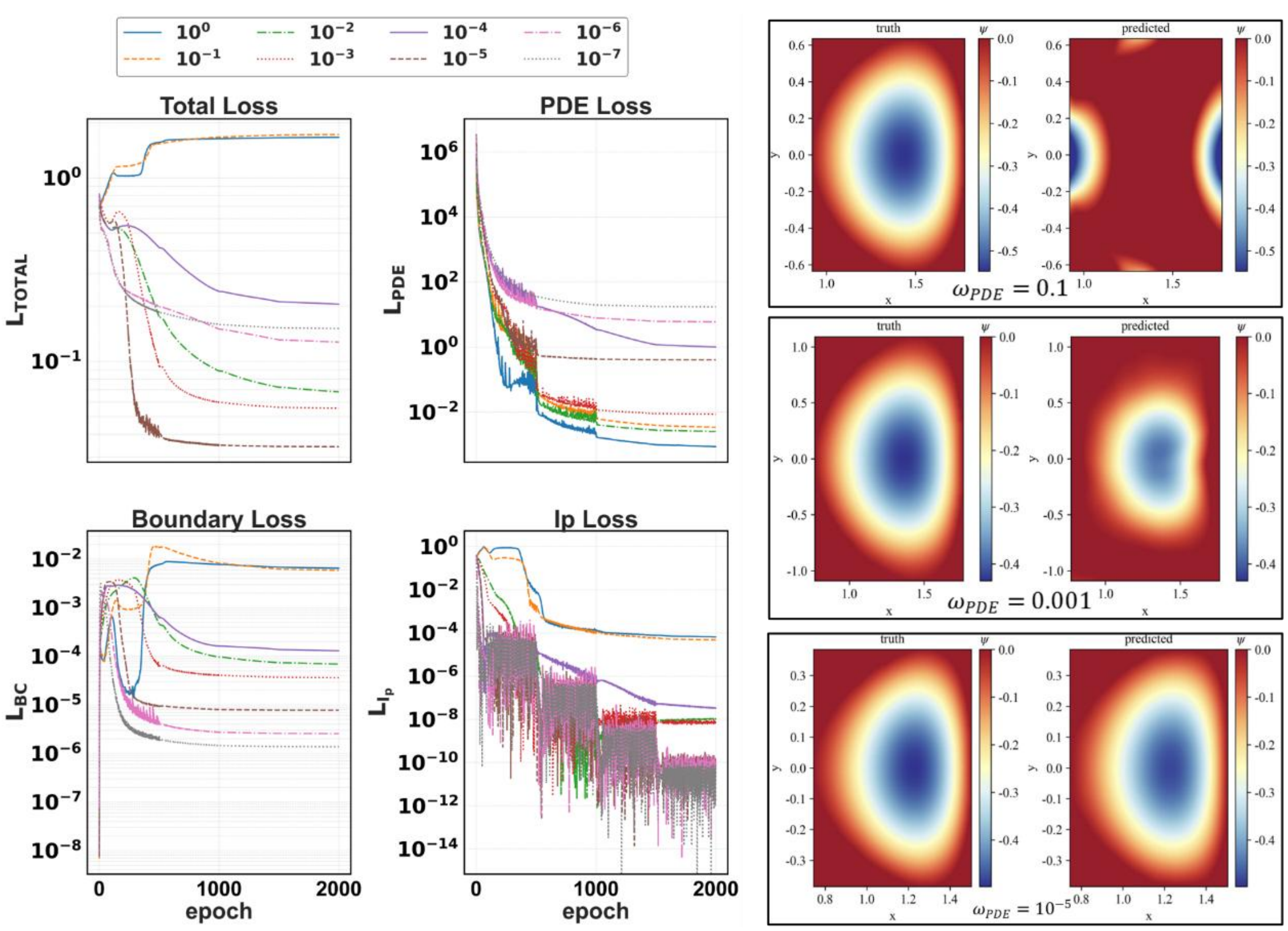

Figure 11. PINO training comparison of a single case (the PDE loss weight varies between 1 and $1\times10^{-7}$, with both boundary and $I_p$ loss weights being 1).

With the optimized loss weights, we trained a parameterized PINO and compared its performance against the supervised learning model. The results of physical metrics in the test dataset are summarized in the box plots in Figure 12. Compared to the supervised TKNO model, the PINO exhibits slightly lower magnetic flux prediction accuracy ($\psi$) compared to supervised learning, propagating into reduced accuracy for derived quantities: internal inductance ($l_i$), edge safety factor ($q_{95}$), and poloidal beta ($\beta_p$). However, the PINO achieved a dramatic reduction in the PDE residual, with the mean error decreasing by nearly four orders of magnitude. Meanwhile, the prediction of the total plasma current ($I_p$) was more accurate. This demonstrates that the PINO architecture successfully prioritizes physical self-consistency over a perfect fit to the training data.

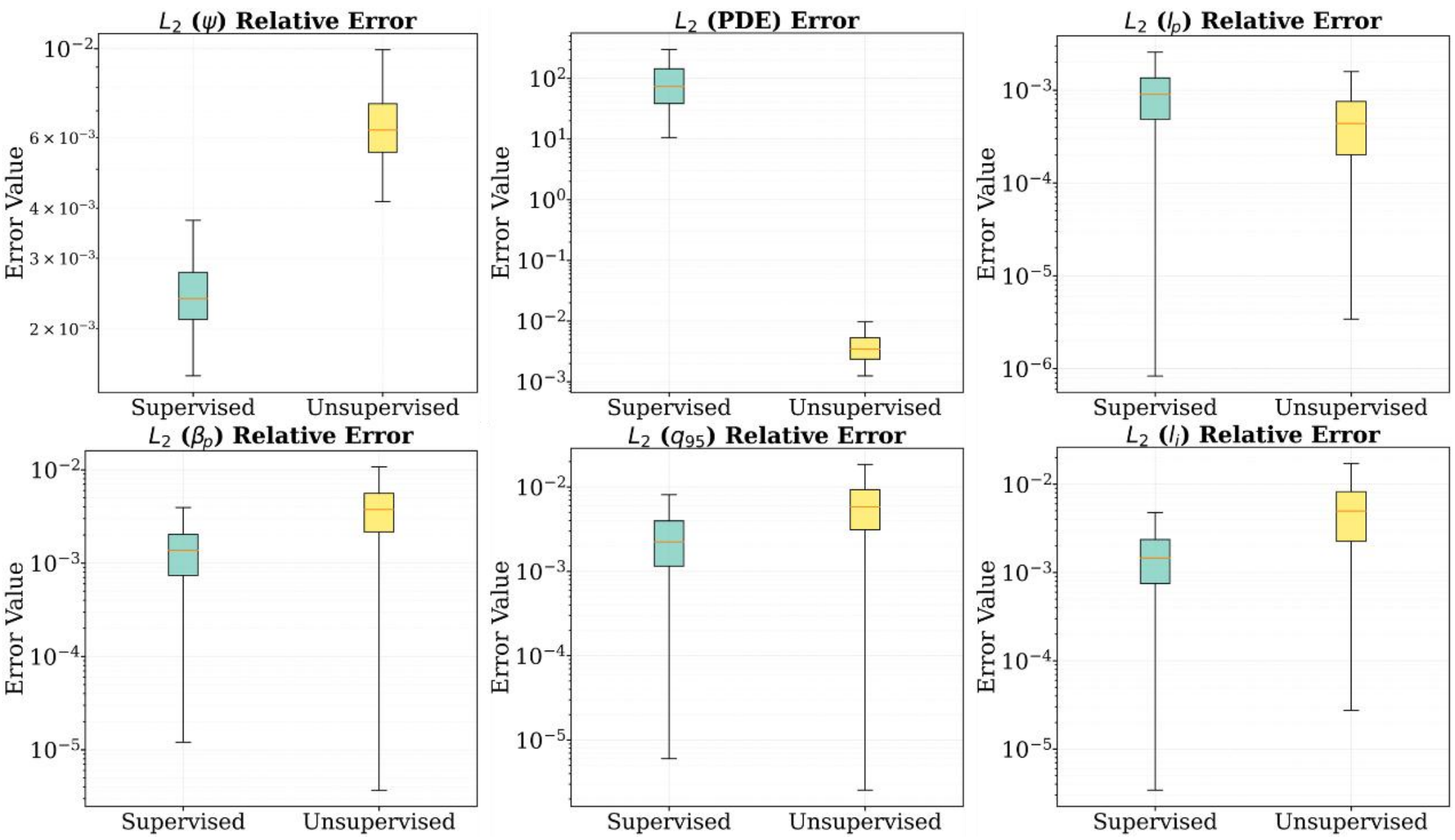

Figure 12. Box plots comparison of physical metrics for the unsupervised TKNO versus the supervised TKNO in the test dataset. The metrics are ordered from left to right, top to bottom: magnetic flux ($\psi$), PDE residual, plasma current ($I_p$), internal inductance ($l_i$), edge safety factor ($q_{95}$), and poloidal beta ($\beta_p$).

Figure 13 provides a visual comparison for a specific test case. While the maximum relative error of the unsupervised-predicted $\psi$ field is larger than that of the supervised model, its error distribution is notably smoother and more continuous. In contrast, the error of the supervised model is scattered and random, further illustrating the PINO's superior physical consistency. For both models, the largest errors are concentrated near the plasma boundary, where gradients are steepest, suggesting that finer grid resolution or hard boundary constraints could lead to further improvements.

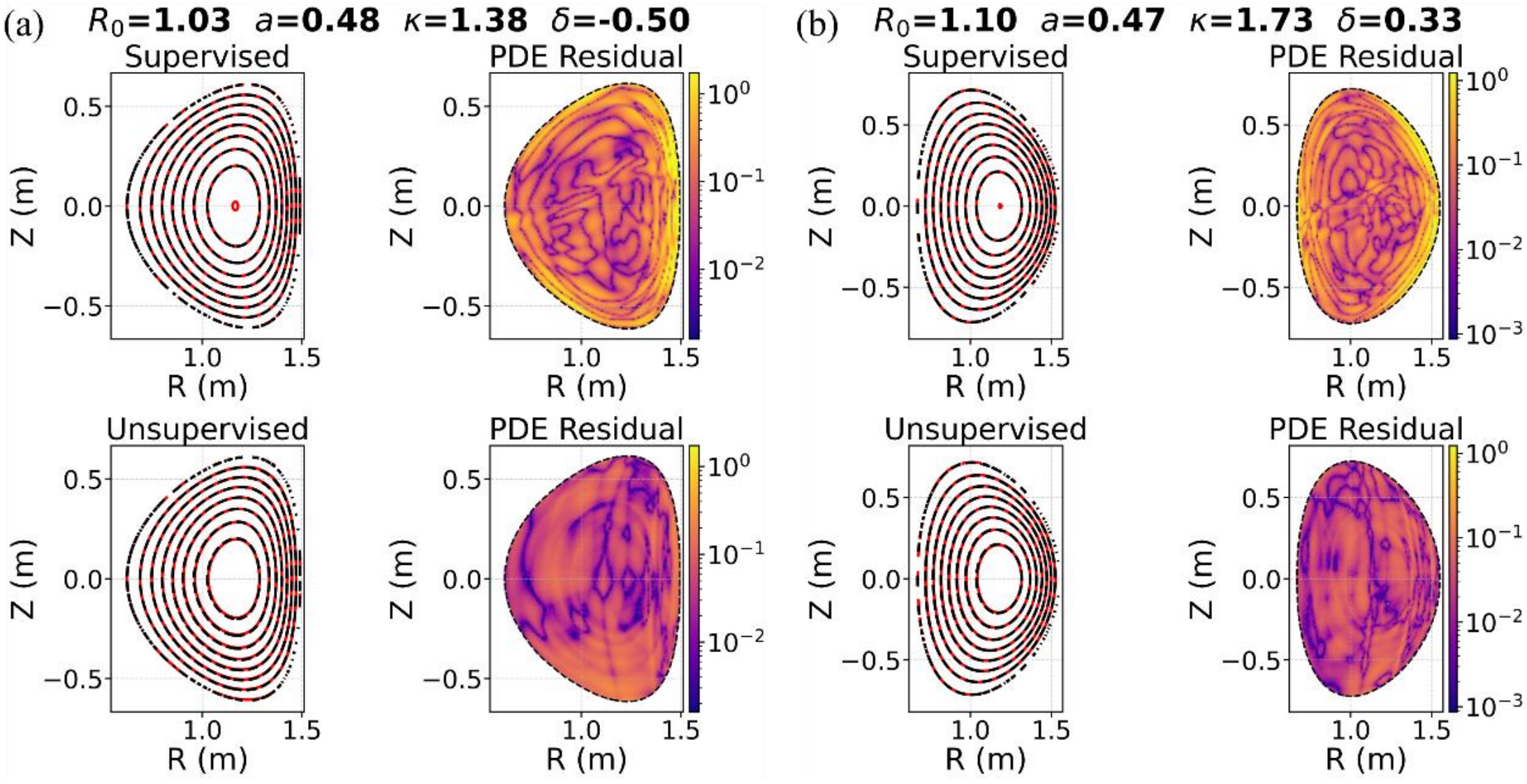

Figure 13. The prediction results of parametric supervised and unsupervised model in two certain test cases.

To prove the validation of FD method, we conducted a preliminary comparison on MLP and KAN architectures as trunk net with the same parameters between PINOs using FD and AD. Table 5 showed that models with FD have better prediction accuracy, this is attributed to the difference gradient values through AD and FD. This also indicates that FD is a viable and computationally efficient alternative for constructing PINOs, particularly when AD is difficult to implement or computationally prohibitive.

Table 5. Comparison of accuracy between AD and FD

| | AD | | | FD | | |
|---|---|---|---|---|---|---|
| model | $L_1(\psi)$ | $L_2(\psi)$ | $L_1$ (PDE) | $L_1(\psi)$ | $L_2(\psi)$ | $L_1$ (PDE) |
| MLP | 0.326 | 0.296 | 15.365 | 0.112 | 0.107 | 8.354 |
| KAN | 0.181 | 0.171 | 6.455 | 0.069 | 0.060 | 4.287 |

### 3.3 Semi-Supervised PINO with Sparse Data

While the unsupervised PINO ensures physical consistency, its predictive accuracy for the magnetic flux field is slightly lower than that of the best supervised models due to challenges in optimizer convergence. Therefore, we investigated a hybrid, semi-supervised learning paradigms by incorporating sparse labeled data into the PINO training process.

This approach supplements the FDM-based PDE loss with a data loss term. For each of the 6,000 samples in our dataset, we randomly sampled a specific number of points from its corresponding computational grid to serve as this labeled data. We then analyzed the impact of varying this number of points per sample on model performance.

The results, presented in Table 6, demonstrate the powerful effect of this hybrid approach and reveal a clear trade-off between predictive accuracy and physical consistency. The analysis reveals an interesting behavior. The unsupervised model (0 points) serves as the $L_2$ ($\psi$) baseline at $9.74\times10^{-3}$. Adding just 10 training points per sample achieves a "win-win": it improves predictive accuracy by 22.9% (to $7.51\times10^{-3}$) and also improves physical consistency, reducing the $L_2$ (PDE) residual by 15.6% (to $7.53\times10^{-3}$).This complex trend in $L_2$ (PDE) (worsening at 1 point, improving at 10 points, then degrading again) suggests that a sparse set of 10 points acts as a synergistic anchor, guiding the optimizer to a better physical solution, whereas too few (1) or too many (100+) points introduce optimization conflict or data-driven trade-offs.

Table 6. Performance on test dataset with varying Number of Monitoring Points

| Point Number (per sample) | $L_2$ ($\psi$) Relative error | Accuracy Improvement (vs. 0 points) | $L_2$ (PDE) Error | Physics Improvement (vs. 0 points) |
|---|---|---|---|---|
| 0 | $9.74\times10^{-3}$ | 0% (Baseline) | $8.92\times10^{-3}$ | 0% (Baseline) |
| 1 | $1.03\times10^{-2}$ | -5.7% | $9.42\times10^{-3}$ | -5.6% |
| 10 | $7.51\times10^{-3}$ | +22.9% | **$7.53\times10^{-3}$** | **+15.6%** |
| 100 | $4.32\times10^{-3}$ | +39.4% | $1.07\times10^{-2}$ | -20.0% |
| 1000 | **$4.22\times10^{-3}$** | **+46.4%** | $1.65\times10^{-2}$ | -85.0% |

However, increasing the data to 100 points per sample yields a significantly larger accuracy boost, cutting the $L_2$ ($\psi$) down to $4.32\times10^{-3}$—a total reduction of 55.6% from the baseline. This substantial accuracy gain comes at the cost of a mild trade-off, with the $L_2$ (PDE) residual ($1.07\times10^{-2}$) rising slightly higher than the baseline. This trade-off becomes sharp after 100 points. Increasing the data tenfold again (to 1,000 points) provides only a marginal 2.3% additional accuracy improvement, while the $L_2$ (PDE) loss degrades significantly (to $1.65\times10^{-2}$). As visualized in Figure 14 (where the 100-point curve closely matches the ground truth), the 100-points-per-sample case represents the “knee” of the curve—the optimal balance. It captures the majority of potential accuracy gains before the returns diminish sharply and the physical consistency trade-off becomes unfavorable.

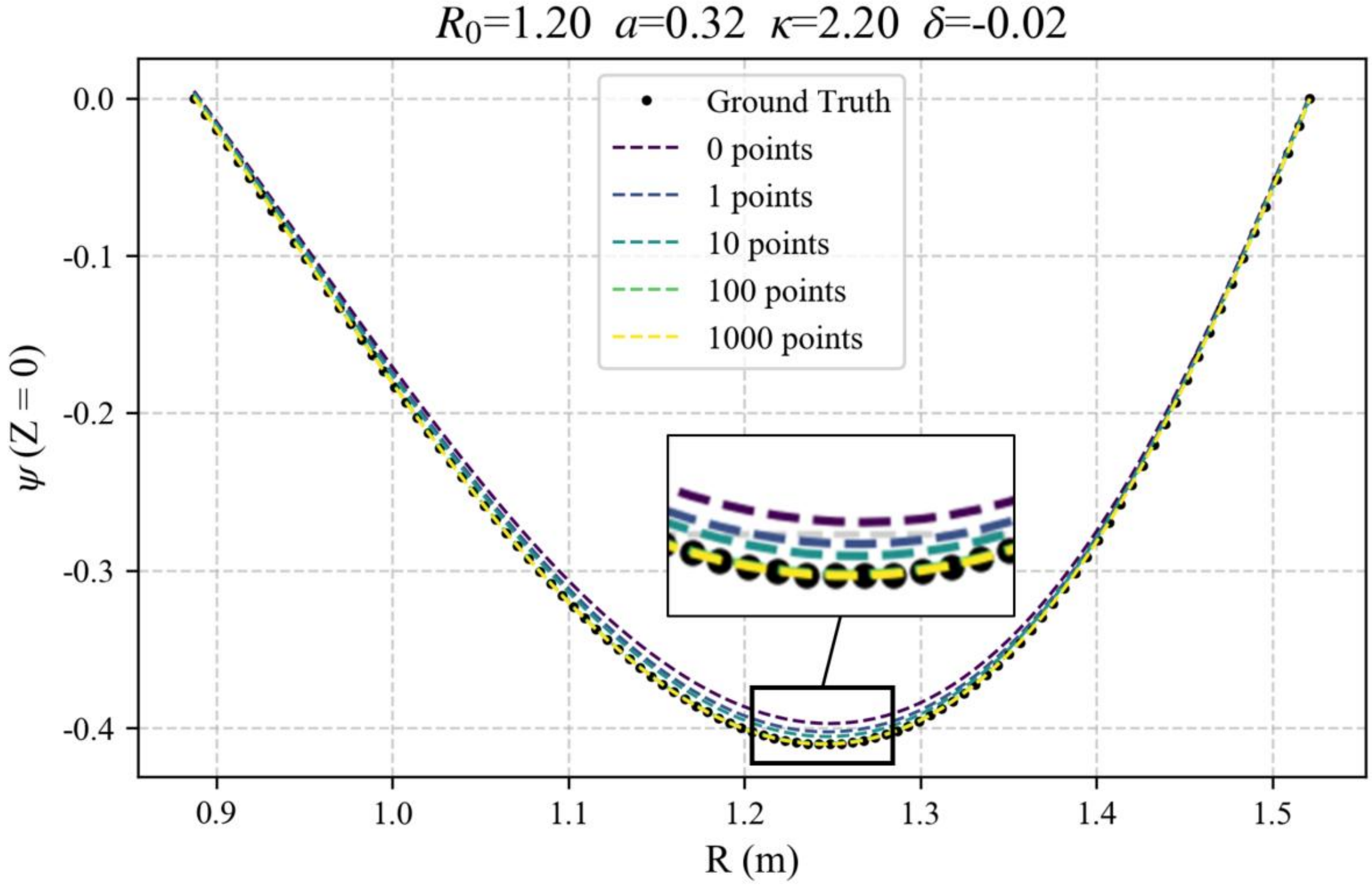

Figure 14. Comparison of magnetic flux profiles predicted by the semi-supervised model.

Ultimately, these findings confirm that a semi-supervised PINO is a highly effective strategy. It successfully merges the high accuracy of data-driven methods with the physical robustness of PINNs, providing a practical and powerful framework for developing surrogate models, particularly in scenarios where measurement data is sparse but valuable.

### 3.4 Evaluation of Extrapolation Performance

For surrogate models intended for real-world applications, evaluating extrapolation performance is the ultimate test of reliability. While generalization measures performance within known parameter regimes, extrapolation tests the model's robustness under unknown operating conditions—a critical requirement for exploring new plasma scenarios.

In this section, we systematically assess the three models’ performance including unsupervised TKNO, supervised TKNO and semi-supervised TKNO (trained with 100 sparse points). The models are evaluated on the OOD dataset ($N$ = 2000), where shape parameters intentionally exceed the training ranges.

The global performance on the OOD dataset is summarized in Figure 15 and Table 7. As anticipated, all models experience performance degradation when moving from the interpolation regime to the extrapolation regime; however, the nature of this degradation reveals fundamental differences in model robustness.

Figure 15 presents the boxplot distribution of the mean $L_2$ relative error for $\psi$. The Supervised model (yellow), while highly accurate in interpolation (4 parameters in training ranges), exhibits "catastrophic failure" in extrapolation. Its error distribution spreads significantly, characterized by a long upper tail and numerous outliers, indicating high unpredictability. Relatively, the unsupervised model (blue) maintains a more compact interquartile range, suggesting that physics constraints successfully bound the error, even if the median accuracy is lower. The semi-supervised model (green) demonstrates the optimal synergy: it achieves a lower median error than the supervised model while maintaining the tight, stable error distribution of the physics-informed approach.

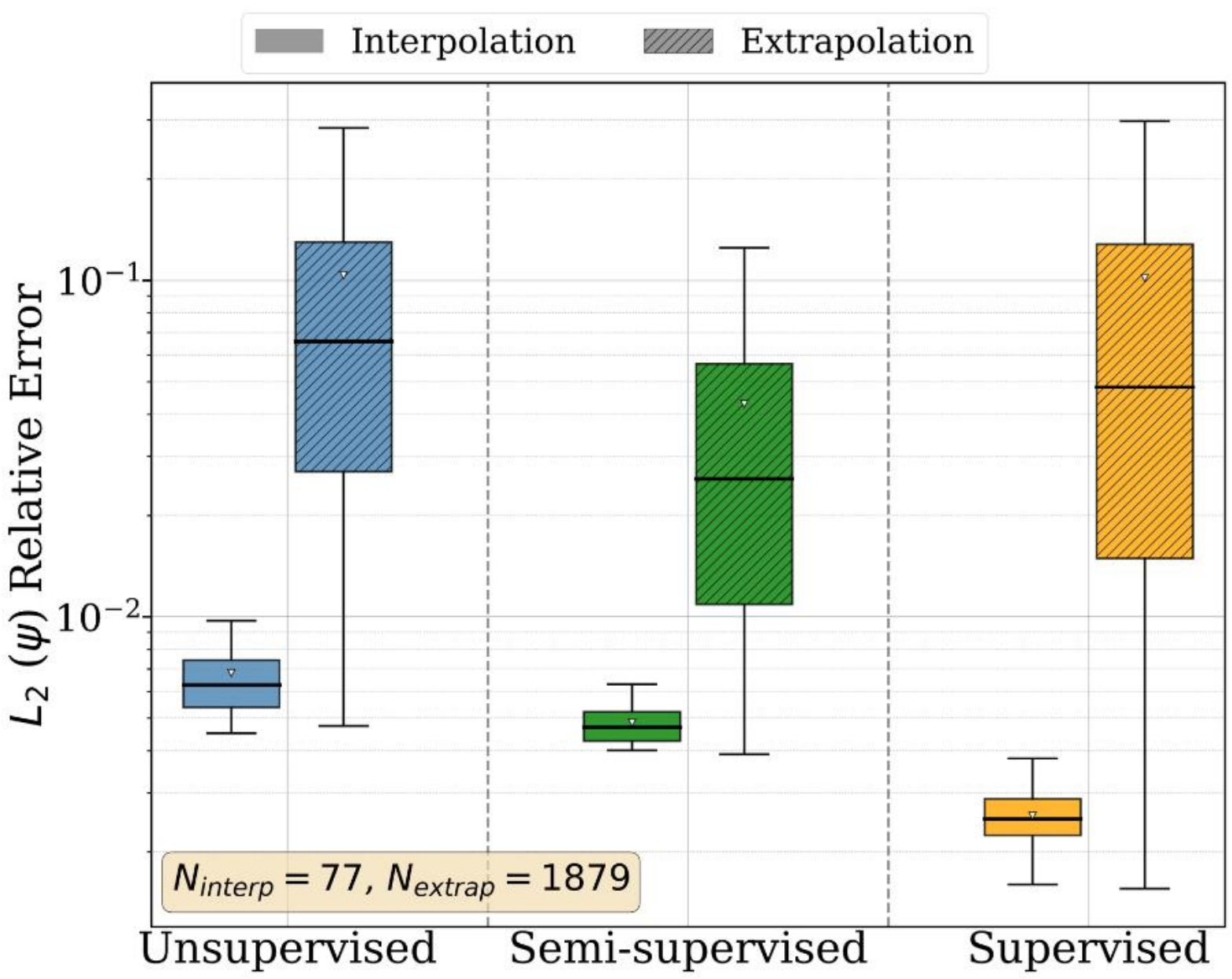

Figure 15. Global performance evaluation: Distribution of the mean $L_2$ relative error for predicted magnetic flux ($\psi$) in interpolation versus extrapolation regimes.

Table 7 quantifies this behavior via the "error growth rate" (the ratio of extrapolation error to interpolation error). The supervised model suffers a massive degradation factor of 39.8×, confirming that pure data-fitting generalizes poorly to new physics. Conversely, the semi-supervised model exhibits "graceful degradation" with the lowest error growth rate of 8.9×. It achieves a mean extrapolation error of 4.76%, significantly outperforming the unsupervised baseline (10.37%) and the unstable supervised model (10.17%).

Table 7. Performance on OOD datasets for three training paradigms

| | Unsupervised | Semi-supervised | Supervised |
|---|---|---|---|
| Interpolation | 0.68% ± 0.21% | 0.48% ± 0.07% | **0.26% ± 0.05%** |
| Extrapolation | 10.37% ± 12.28% | **4.76% ± 4.77%** | 10.17% ± 14.36% |
| Error growth rate | 15.2 | **8.9** | 39.8 |

To understand the drivers of extrapolation error, Figure 16 dissects performance across the four individual shape parameters ($R_0$, $a$, $\kappa$, $\delta$). In these subplots, the gray shaded regions represent the training distribution, while the white regions represent extrapolation zones. (Note: For any data point shown, the specific parameter is fixed within the interval while others vary according to the LHS distribution).

The instability of supervised models and robustness of semi-supervised model emerge across all dimensions. As parameters move further from the training range (e.g., $a > 0.5$ or $\delta < -0.5$), the supervised model frequently exhibits large error spikes (long whiskers), confirming its inability to capture the underlying governing laws.

In contrast the semi-supervised model maintain stability. Notably, in challenging regimes such as the high minor radius interval ( $a \in [0.5, 0.7]$ ), it remains an error close to that within the training range. This confirms that sparse data combined with physics constraints is more effective than dense data alone when exploring novel geometries.

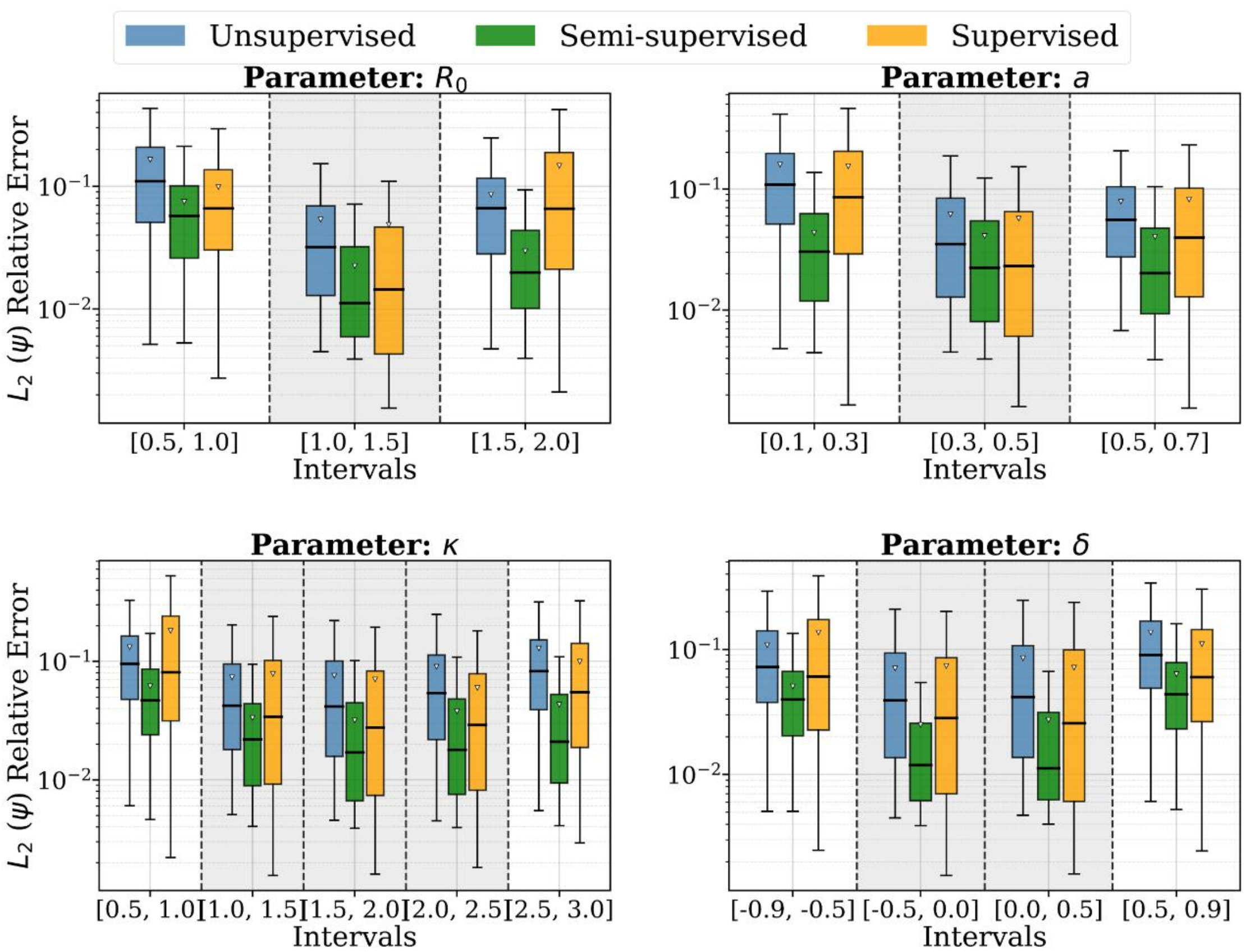

Figure 16. Parameter sensitivity analysis: Dependence of prediction error on individual geometric shape parameters ($R_0$, $a$, $\kappa$, $\delta$) in the extrapolation regime.

To visualize the physical fidelity of the predictions, Figure 17 presents two extreme extrapolation cases. Case (a) represents a “stress test” with significant negative triangularity ($\delta$ = -0.8). The supervised model fails completely, predicting a severely distorted core with non-physical artifacts. In stark contrast, the semi-supervised model successfully reconstructs the D-shaped topology, closely matching the Ground Truth. This proves that the embedded PDE constraints prevent the model from collapsing into non-physical solutions. Case (b) illustrates a high-elongation, high-triangularity boundary case ($\kappa$ = 2.7, $\delta$ = 0.7). While the Supervised model produces a reasonable approximation, its solution contains high-frequency noise and irregular contours. The Semi-Supervised result is notably smoother and maintains strict physical consistency, validating its potential for reliable use in next-generation tokamak control systems.

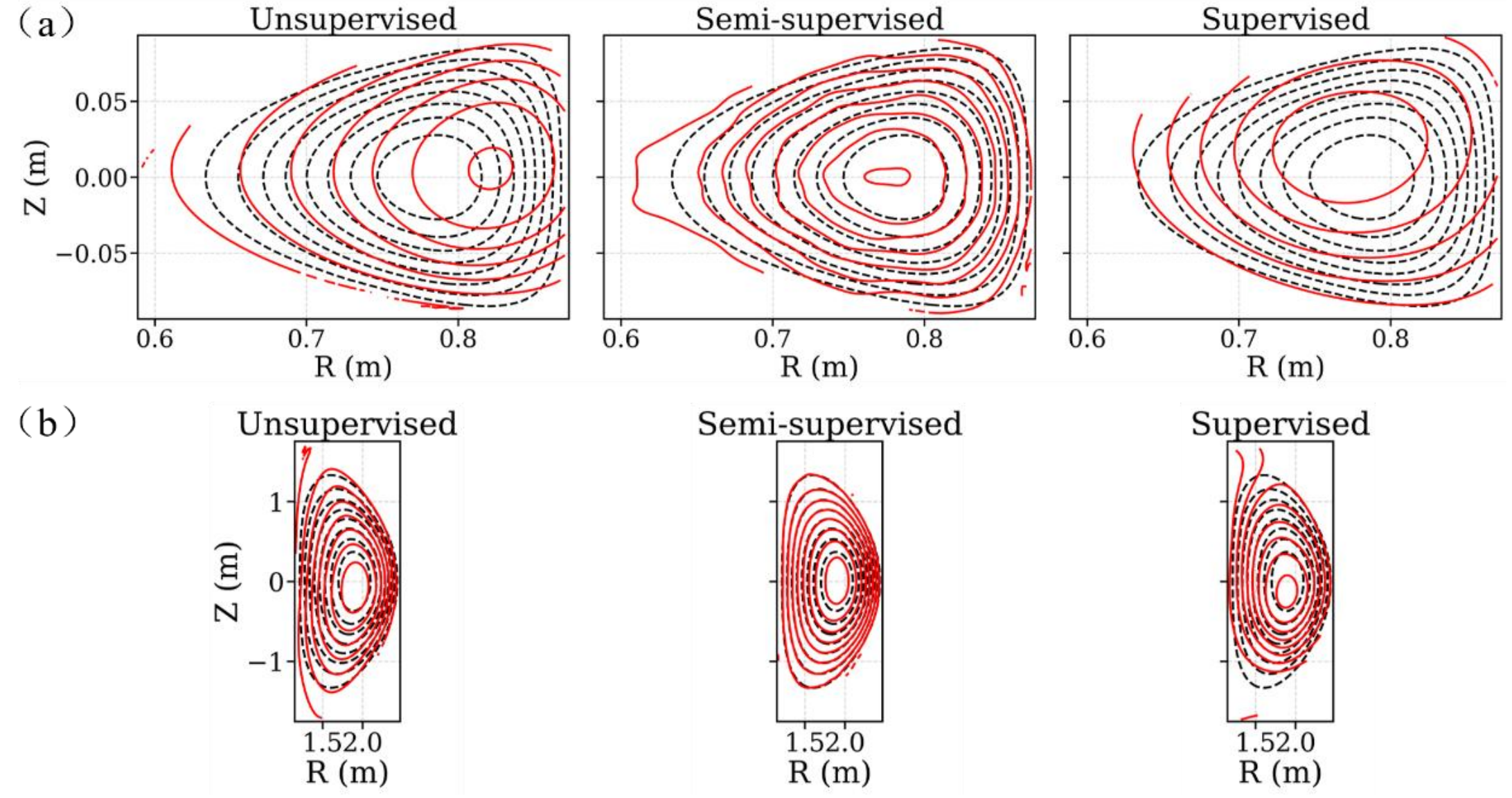

Figure 17. Comparison of the predicted magnetic flux ($\psi$) from the three models against the Ground Truth in extreme extrapolation scenarios. (a) $R_0$ = 0.7, $a$ = 0.1, $\kappa$ = 0.7, $\delta$ = -0.8. (b) $R_0$ = 1.8, $a$ = 0.7, $\kappa$= 2.7, $\delta$= 0.7.

### 3.5 Inference Speed Comparison

All benchmarks were conducted on a workstation featuring an Intel Xeon Gold 6530 CPU (112 cores) and an NVIDIA RTX 5090 GPU. The MATLAB-based FDM-SOR baseline was restricted to CPU execution due to the serial nature of its iterative solver. Neural networks were evaluated on the CPU (PyTorch) and GPU (PyTorch, TorchScript, and TensorRT). To ensure statistical reliability, average inference times for single case were calculated from 1,000 runs following a 10-iteration warmup to eliminate initialization overhead.

As summarized in Table 8. Comparison of the inference time for different methods., the FDM-SOR method requires 10.2 s for a single inference on the CPU. In contrast, TensorRT optimization dramatically reduces neural network latency to the millisecond range. Specifically, the T-K model, which offers the optimal $\psi$ prediction accuracy, achieves an inference latency of 1.16 ms. This rapid processing speed successfully meets the rigorous requirements for kilohertz-rate feedback control.

Table 8. Comparison of the inference time for different methods.

| Method | CPU | PyTorch | TorchScript | TensorRT |
|---|---|---|---|---|
| FDM-SOR | 10.2s | - | - | - |
| M-K | 0.4s | 0.42 ms | 0.44 ms | **0.30 ms** |
| K-K | 4.3s | 13.31 ms | 2.59 ms | **0.69 ms** |
| F-K | 3.9s | 6.49 ms | 4.72 ms | - |
| C-K | 4.6s | 4.56 ms | 3.86 ms | **2.11 ms** |
| T-K | 5.9s | 14.29 ms | 6.25 ms | **1.16 ms** |

## 4 SUMMARY and CONCLUSIONS

As machine learning algorithm plays an increasingly vital role in nuclear fusion research and development, this study systematically evaluated PINO learning for solving the fixed-boundary Grad-Shafranov equation with realistic nonlinear GAQ current profiles. Through comprehensive benchmarking of five neural operator architectures and rigorous comparison of supervised, unsupervised, and semi-supervised learning paradigms, we establish that semi-supervised PINO achieves the optimal balance between predictive accuracy, physical consistency, and extrapolation robustness required for next-generation fusion reactor control systems.

**Key Findings**:

**1) neural operator enables parameterized equilibrium analysis at unprecedented speed**: the novel Transformer-KAN Neural Operator (TKNO) achieved state-of-the-art accuracy (0.25% ± 0.05% mean $L_2$ relative error) under supervised training. However, all data-driven models exhibited large PDE residuals (on the order of $10^2$) and catastrophic extrapolation failure (39.8× error increase), demonstrating that high data-fitting accuracy does not guarantee physical consistency.

**2) physics-informed training dramatically improved physical fidelity**: incorporating PDE residual, boundary conditions, and plasma current constraints reduced PDE residuals by nearly four orders of magnitude (from $\sim 10^2$ to $\sim 10^{-2}$), though at the cost of slightly lower flux prediction accuracy. The unsupervised approach demonstrated predictable, bounded extrapolation behavior (15.2× error increase)—critical for safe deployment in operational tokamaks.

**3) semi-supervised learning achieved optimal synergy by integrating sparse labeled data** (100 interior points) with physics constraints, recovering data-driven accuracy levels (0.48% interpolation error) while maintaining low PDE residuals ($\sim 10^{-2}$). Most critically, extrapolation analysis on 2000 out-of-distribution samples revealed “graceful degradation” with the smallest error increase factor (8.9×) and mean extrapolation error of 4.76% ± 4.77% across extreme parameter regimes.

4) **TensorRT acceleration achieved millisecond-level real-time inference in nuclear fusion experiments**, representing over three-order-of-magnitude speedup compared to traditional solvers (>10 seconds) and meeting kilohertz-rate feedback control requirements.

**Limitations**. Our fixed-boundary assumption simplifies real experimental scenarios requiring free-boundary solvers; GAQ polynomial profiles do not capture all profile complexities (reversed-shear, current holes); 2D axisymmetric geometry neglects 3D effects critical for advanced scenarios; and extrapolation to entirely novel physics (different profile families) remains unvalidated. We hope this preliminary work will ignite interest in the application of rigorous PIML for experiments and simulations in nuclear fusion.

**Future directions**. Priority extensions include: (1) free-boundary equilibrium coupling to vacuum regions and coil circuits for experimental control integration; (2) inverse equilibrium reconstruction from magnetic diagnostics with uncertainty quantification; (3) time-dependent evolution modeling for transient prediction; (4) integration with whole-device frameworks (OMFIT, TRANSP) for multi-physics optimization; and (5) deployment as simulation environments for reinforcement learning-based autonomous plasma control—the frontier demonstrated by recent breakthroughs in instability avoidance. Beyond equilibrium, the semi-supervised paradigm established here—combining sparse data guidance with rigorous physics constraints for graceful extrapolation—offers a generalizable framework applicable to other fusion-relevant PDE, proving essential for realizing commercial fusion energy.

**Acknowledgement**

We are particularly grateful to ENN for the scientific research funding supporting this study.